\documentclass[final,3p,times]{elsarticle}
\usepackage{amssymb}
\usepackage{amsmath}
\usepackage{graphicx}
\journal{Physica A}

\def\Eq#1{Eq.~(\ref{#1})}
\def\Eqs#1{Eqs.~(\ref{#1})}
\def\Fig#1{Fig.~\ref{#1}}	

\def\no{\nonumber\\}

\def\>{\rangle}
\def\<{\langle}

\def\half{\frac{1}{2}}
\def\dt#1{\frac{d{#1}}{dt}}
\def\adg{a^\dagger}

\def\dg{\dagger}

\def\lam{\lambda}
\def\kap{\kappa}
\def\w{\omega}

\def\del{\delta}

\def\v#1{\mbox{\boldmath$#1$}}

\def\thv{\mbox{\boldmath$\theta$}}
\def\thh{\mbox{$\hat{\theta}$}}
\def\thvh{\mbox{\boldmath$\hat{\theta}$}}
\def\thv{\mbox{\boldmath$\theta$}}
\def\av{\mbox{\boldmath$a$}}
\def\bv{\mbox{\boldmath$b$}}
\def\cv{\mbox{\boldmath$c$}}
\def\ev{\mbox{\boldmath$\eta$}}
\def\e{\mbox{\boldmath$e$}}
\def\gv{\mbox{\boldmath$g$}}
\def\g{g}

\def\Pv{\mbox{\boldmath$\Pi$}}
\def\Sv{\mbox{\boldmath$\Gamma$}}
\def\Thv{\mbox{\boldmath$\Theta$}}
\def\Sigv{\mbox{\boldmath$\Sigma$}}

\def\gam{\gamma}
\def\sig{\sigma}
\def\Gam{\Gamma}
\def\Del{\Delta}
\def\al{\alpha}
\def\bt{\beta}
\def\th{\theta}

\def\Km{K}
\def\Ku{{\underline{K}}}
\def\X{\left(\!\!\begin{array}{c} 1\\0 \end{array}\!\!\right)}

\def\xt{\widetilde{x}}

\def\d{\partial}
\def\T{\text{T}}
\def\u#1{\underline{#1}}
\def\KL{\text{KL}}
\def\CL{\text{CL}}
\def\HPZ{\text{HPZ}}

\begin{document}

\begin{frontmatter}

\title{Solutions of generic bilinear master equations for a quantum oscillator -- positive and factorized conditions on stationary states}

%\author{B.~A.~Tay }
%\email{BuangAnn.Tay@nottingham.edu.my}
%\affiliation{Foundation Studies, Faculty of Engineering, The University of Nottingham Malaysia Campus, Jalan Broga, 43500 Semenyih, Selangor, Malaysia}

%\date{\today}

\author[1]{B. A. Tay}
\ead{BuangAnn.Tay@nottingham.edu.my}
\address[1]{Foundation Studies, Faculty of Engineering, The University of Nottingham Malaysia Campus, Jalan Broga, 43500 Semenyih, Selangor, Malaysia}

\date{\today}

\begin{abstract}
We obtain the solutions of the generic bilinear master equation for a quantum oscillator with constant coefficients in the Gaussian form. The well-behavedness and positive semidefiniteness of the stationary states could be characterized by a three-dimensional Minkowski vector. By requiring the stationary states to satisfy a factorized condition, we obtain a generic class of master equations that includes the well-known ones and their generalizations, some of which are completely positive. A further subset of the master equations with the Gibbs states as stationary states is also obtained. For master equations with not completely positive generators, an analysis on the stationary states for a given initial state isuggests conditions on the coefficients of the master equations that generate positive evolution.
\end{abstract}


\end{frontmatter}

%\maketitle

%%%%%%%%%%%%%%%%%%%%%%%%%%
%       Sectionn         %
%%%%%%%%%%%%%%%%%%%%%%%%%%
\section{Introduction}

When we obtain the master equations of a system in contact with an environment, there are often assumptions made to facilitate the derivation of the effective influence of the environment on the system \cite{Breuer,Gardiner}. For instance, we assume that the initial density matrix of the system and environment is factorized, the coupling between the system and its environment is weak, the rotating wave approximation is valid, the memory effects are negligible, or the generators of the master equation should be completely positive, and etc. Appropriate derivations should produce master equations with well-behaved solutions.

However, anomaly could arise if we carelessly apply the master equations to situations that are not consistent with the assumptions made in their derivations, such as studying low temperature behavior of a system with the Caldeira-Leggett equation \cite{Haake85,Diosi93a,Tameshtit12}, obtaining an Markovian equation by ignoring the memory effects or using the rotating wave approximation \cite{Suarez92,Munro96,Whitney08,Fleming10}, or starting with initial conditions inconsistent with the factorized assumptions \cite{Pechukas94,Shaji04}.

An issue that has received wide attention is the completely positive \cite{Kossa76,Lindblad76} nature of the generators of the master equations. Initial states that are positive could evolve outside their positive domain during some interval of the time evolution when the master equations do not have completely positive generators. However, there is no a priori reason why master equations must be completely positive \cite{Pechukas94}. Not completely positive generators with appropriately chosen coefficients could maintain the positivity of the states for a given initial state for all time \cite{Shaji05}. The systems are then properly behaved as far as positivity is concerned.

It is in this spirit that we start with a generic master equation that satisfies the essential hermitian and trace preserving requirements of a reduced dynamics \cite{Sudarshan61,Talkner81,Tay16}. We obtain its explicit solutions in Gaussian form and analyzed the stationary states. We further identify a factorized condition on the stationary states that is strong enough to produce a generic class of master equations. This class includes the well-known master equations as special cases, i.e., the Kossakowski-Lindblad (KL) equation for quantum optical systems, the Caldeira-Leggett equation (CL) and the Hu-Paz-Zhang (HPZ) equation for quantum Brownian motion, as well as their generalizations, some of which are found to be completely positive. In this respect, Ref.~\cite{Vacchini02} discussed the role of the Gibbs states, which are examples of states satisfying the factorized condition, and other requirements in deciding the form of quantum master equations. By analyzing the positivity of the stationary states of the master equation, we could infer the coefficients of the master equation that yield positive evolution for a given initial state.

Solutions to various master equations had been obtained, such as for a family of equations whose forms are closely related to the Fokker-Planck equations \cite{Agarwal71,Walls85}, the KL equation \cite{Walls08}, the HPZ equation \cite{Hu11}, the generic equations \cite{Tameshtit13} that include the quantum Brownian motion, as well as solutions in terms of the second moments \cite{Isar99}. The uncertainty principle \cite{Dekker84,Vacchini02} and the Schwartz inequality \cite{Talkner81} were invoked to give constraints on the second moments of the observables of the equations. Numerical solutions could be obtained using the Gaussian ansatz \cite{Giulini}.

Our method is closely related to Ref.~\cite{Tameshtit13}, where a different basis of generators was used to derive the general solutions to master equations for quantum oscillators with time-dependent coefficients. Generic conditions on the positivity of the equations were also given. However, owing to the complexity of the exact time-dependent solutions, the effects of the coefficients on the behaviours of the solutions are not transparent, and it is difficult to analyze them. Therefore, an analysis on the solutions to the generic master equation with constant coefficients could provide insights on the structure of the master equations and their roles in deciding the positivity of the time evolution.

We first provide the most general form of the master equations in Section \ref{SecGenTmEvo}, and discuss the necessary and sufficient conditions for positive semidefinite density matrices in the Gaussian form. We then obtain the four-dimensional (4D) matrix representation of the time evolution operator in Section \ref{Sec4DU}, and decompose it into a Baker-Campbell-Hausdorff (BCH) formula in Section \ref{SecBCHUt}. The solutions to the generic master equation are then obtained in Section \ref{SecSolMME}. This is followed by a study on the properties of the generic stationary states in Section \ref{SecBehavSol}. In Section \ref{SecMEsep}, we make use of the factorized condition to obtain the known master equations and their generalizations. This is followed by discussions in Section \ref{SecDisc}. Some of the details of the calculations are presented in the appendices.

%%%%%%%%%%%%%%%%%%%%%%%%%%
%       Sectionn         %
%%%%%%%%%%%%%%%%%%%%%%%%%%
\section{The generic master equation}
\label{SecGenTmEvo}

%%%%%%%%%%%%%%%%%%%%%%%%%%
%       Sectionn         %
%%%%%%%%%%%%%%%%%%%%%%%%%%
\subsection{Generator of time evolution}
\label{SecTmEvo}

The density matrices of quantum oscillators evolve according to the equation
%%%
\begin{align}   \label{rhot}
    \frac{\d \rho}{\d t}=-\Km\rho\,,
\end{align}
%%%
where we separate the generator of time evolution $\Km$ into two parts
%%%
\begin{subequations}
\begin{align}   \label{K}
    \Km&=\Km_0+\Km_1\,,\\
    \Km_0&\equiv\th_0 iL_0 +\th_1 iM_1 +\th_2 iM_2\,, \label{K0}\\
    \Km_1&\equiv \gam(O_0-1/2) +\eta_0O_+ +\eta_1L_{1+} +\eta_2L_{2+}\,, \label{K1}
\end{align}
\end{subequations}
%%%
in which $\gam$, $\th_i, \eta_i, i=0, 1, 2$ are real coefficients. $K$ is the most general form of the generators that preserves the hermiticity of density matrices and conserves the probability of reduced dynamics \cite{Tay16}. We consider only constant coefficients here. In the position coordinates, the operators are
%%%
\begin{subequations}
\begin{align} \label{iL0}
    iL_0 = \frac{i}{2}\left(-\frac{\d^2}{\d Q\d r}+Qr\right)\,,\,\,
    iM_1= \frac{i}{2}\left(\frac{\d^2}{\d Q\d r}+Qr\right)\,,\,\,
    iM_2= -\frac{1}{2}\left(Q\frac{\d}{\d Q}+r\frac{\d}{\d r}+1\right)\,,\,\,
    O_0 = -\frac{1}{2}\left(Q\frac{\d}{\d Q}-r\frac{\d}{\d r}\right)\,,
\end{align}
%%%
%%%
\begin{align}   \label{O+}
    O_+ = \frac{1}{4}\left(\frac{\d^2}{\d Q^2}-r^2\right)\,, \qquad
    L_{1+} = -\frac{1}{4}\left(\frac{\d^2}{\d Q^2}+r^2\right)\,, \qquad
    L_{2+}= -\frac{i}{2}r\frac{\d}{\d Q}\,,
\end{align}
\end{subequations}
%%%
where the center and relative coordinates are defined as
%%%
\begin{align} \label{Qrxx}
    Q&\equiv\frac{1}{2}(x+\xt)\,,   \qquad r\equiv x-\xt\,,
\end{align}
%%%
respectively, in which $\xt$ and $x$ are the position coordinates in the bra- $|\xt\>$ and ket-space $\<x|$, respectively. The commutation relations between these generators can be found in Ref.~\cite{Tay16}.

Let us now discuss the effect of each generator. $iL_0$ is the free Liouvillean of the harmonic oscillator. Its coefficient is related to the natural frequency of the oscillator by $\th_0=2\w_0$. The operator $iM_1$ renormalizes or shifts the natural frequency to $\w_\text{renorm}\equiv (\w_0^2-\th_1^2/4)^{1/2}$. This renormalization occurs, for example, when the oscillator is coupled to a field, which in the open quantum system context eventually reduces to the environment.

The $iM_2$ introduces damping to the oscillator. From the solution obtained later in Section \ref{SecGensol}, we find that $iM_2$ modifies the frequency of the damped oscillator to
$\w_\text{damp}\equiv(\w_\text{renorm}^2-\th_2^2/4)^{1/2}$. The relative magnitude of the coefficient of $iM_2$ decides whether the oscillator is underdamped, critically damped, or overdamped, when $\w_\text{renorm}$ is greater than, equal to, or smaller than $\th_2/2$, respectively. In the overdamped situation, $\w_\text{damp}$ turns imaginary. The $O_0-1/2$ term contributes to the relaxation and excitation of the oscillator.

The coefficients of $O_+$ and $L_{1+}$ is a function of the temperature of the environment. The $\d^2/\d Q^2$ term is analogous to the diffusion term in the classical diffusion equation. It originates from the random motion of the environment degrees of freedom. It tends to spread the probability distribution function along the $Q$ or diagonal direction. On the other hand, the $r^2$ term is responsible for destroying the off-diagonal components of the density matrices, or decoherence \cite{Zurek91}.

The generator $L_{2+}$ is a diffusion term that has no classical counterpart in one space dimension. We will find that it leads to negative probability when its effects are dominant over the $O_+$ and $L_{1+}$ terms.

%%%%%%%%%%%%%%%%%%%%%%%%%%
%       Sectionn         %
%%%%%%%%%%%%%%%%%%%%%%%%%%
\subsection{Gaussian ansatz}
\label{SecGaussian}

We consider solutions in the Gaussian form,
%%%
\begin{align}   \label{rhoGaussian}
    \rho(Q, r;t) &=\sqrt{\frac{2\mu(t)}{\pi}}e^{-4\mu(t){Q^2}/{2}-\kap(t)  iQr-[\mu(t)+\nu(t)]{r^2}/{2}}\,,
\end{align}
%%%
where $\mu, \kap$ and $\nu$ are real functions of time \cite{Giulini}. Density matrices of this form have zero first moments. Inserting \Eq{rhoGaussian} into the equation of motion \eqref{rhot}, we find that the coefficients evolve according to the following set of equations,
%%%
\begin{subequations}
\begin{align}   \label{ratemu}
   \dt{\mu}&=(\gam+\th_2)\mu+(\th_0-\th_1)\mu\kap+2(\eta_0-\eta_1)\mu^2\,,\\
   \dt{\kap}&=\half(\th_0+\th_1)+\th_2\kap-\half(\th_0-\th_1)\big[4\mu(\mu+\nu)-\kap^2\big]+2\eta_2\mu+2(\eta_0-\eta_1)\mu\kap\,,\label{ratekap}\\
   \dt{({\mu}+\nu)}&=\half(\eta_0+\eta_1)-(\gam-\th_2)(\mu+\nu)+(\th_0-\th_1)(\mu+\nu)\kap-\eta_2\kap-\half(\eta_0-\eta_1)\kap^2\,.\label{ratemunu}
%   \dt{\xi}&=\half(\gam+\th_2)\xi+2(\th_0-\th_1)\left(\mu\chi+\kap \frac{\xi}{4}\right)+2(\eta_0-\eta_1)\mu\xi\,,\\
%   \dt{\chi}&=-\half(\gam-\th_2)\chi+\half(\th_0-\th_1)[\kap\chi-(\mu+\nu)\xi]+\half\eta_2\xi+\half(\eta_0-\eta_1)\kap\xi\,,\\
%   \dt{h}&=-\half(\gam+\th_2)+\half(\th_0-\th_1)(\xi\chi-\kap)+(\eta_0-\eta_1)\left(\frac{\xi^2}{4}-\mu\right)\,.\label{rateh}
\end{align}
\end{subequations}
%%%
\Eqs{ratemu}-\eqref{ratemunu} are coupled nonlinear equations of $\mu, \kap$ and $\nu$. If we add kernels linear in the position coordinates, i.e.~$\exp[-\xi(t) Q-\chi(t) ir]$, to $\rho$, the evolutions of $\mu, \kap, \nu$ \eqref{ratemu}-\eqref{ratemunu} are unaffected, though the first moments are now nonzero. The kernels linear in the coordinates can be transformed away by displacement operators \cite{Klauder68,Gardiner}. Hence, $\mu, \kap$ and $\nu$ alone are sufficient to determine the behaviours of the dissipative dynamics and without loss of generality we will not consider kernels linear in the coordinates here.

We normalize the density matrices \eqref{rhoGaussian} according to
%%%
\begin{align}   \label{normrho}
    \int^\infty_{-\infty} \rho(Q,0;t) dQ =1\,,
\end{align}
%%%
where
%%%
\begin{align}   \label{PDF}
    \rho(Q,r=0;t)=\sqrt{\frac{2\mu(t)}{\pi}} e^{-2\mu(t)Q^2}
\end{align}
%%%
is a probability distribution function. The necessary and sufficient conditions for the positive semidefiniteness of $\rho$ is \cite{Simon87}
%%%
\begin{align}   \label{nscond}
    \mu(t)>0\,,\qquad \text{and} \qquad \nu(t)\geq 0\,,
\end{align}
%%%
for all time.

Our objective is to find the solutions to the set of equations of motion \eqref{ratemu}-\eqref{ratemunu} that yield positive semidefinite $\rho$.

%%%%%%%%%%%%%%%%%%%%%%%%%%
%       Sectionn         %
%%%%%%%%%%%%%%%%%%%%%%%%%%
\subsection{Second moments}
\label{Sec2ndmoment}

Before we present the solutions to the equations, we divert to study the relationships between the second moments with the coefficients $\mu, \kap, \nu$, and the positive semidefiniteness conditions \eqref{nscond}. The expectation value of an operator $\hat{O}$ is defined as
%%%
\begin{align}   \label{Aexpect}
    \<\hat{O}\>_t=\text{tr}\big[\hat{O}\hat{\rho}(t)\big]
            =\int_{-\infty}^{\infty}O(Q,r)\rho(Q,r;t)\big|_{r=0}dQ\,,
\end{align}
%%%
where the subscript $t$ denotes its time-dependence. In the $Q, r$ coordinates, the position and the momentum operators are respectively given by
%%%
\begin{align}   \label{xpQr}
    x(Q,r)=Q+\frac{r}{2}\,, \qquad p(Q,r)=-i\frac{\d}{\d
    x}=-i\left(\half\frac{\d}{\d Q}+\frac{\d}{\d r}\right)\,,
\end{align}
%%%
where we use the units $\hbar=1$.

For density matrices of the Gaussian form \eqref{rhoGaussian}, the first moment of the position and momentum operators vanish, $\<\hat{x}\>_t=0=\<\hat{p}\>_t$. The second moments are
%%%
\begin{align}   \label{2ndx}
    \<\hat{x}^2\>_t&=\frac{1}{4\mu(t)}\,,\qquad
    \<\hat{p}^2\>_t=\mu(t)+\nu(t)+\frac{\kap^2(t)}{4\mu(t)}\,,\qquad
    \<\hat{x}\hat{p}\>_t=\<\hat{p}\hat{x}\>_t+i=-\frac{\kap(t)}{4\mu(t)}+\frac{i}{2}\,.
\end{align}
%%%
Substituting \Eqs{2ndx} to the Schwartz inequality $\<\hat{x}^2\>_t  \<\hat{p}^2\>_t \geq\<\hat{x}\hat{p}\>_t\<\hat{p}\hat{x}\>_t$, or the generalized uncertainty relation \cite{Dekker84}, $\big[\<\hat{p}^2\>_t-\<\hat{p}\>_t^2\big]\big[\<\hat{x}^2\>_t-\<\hat{x}\>_t^2\big]
-\big[\frac{1}{2}(\<\hat{x}\hat{p}\>_t+\<\hat{p}\hat{x}\>_t)-\<\hat{p}\>_t\<\hat{x}\>_t\big]^2\geq 1/4$, where we use the units $\hbar=1$, the condition $\nu(t)\geq 0$ is reproduced. Together with $\mu>0$, we recover the necessary and sufficient conditions for the positive semidefiniteness of the time evolution \eqref{nscond}. We arrive at the same conclusion when we apply the second moments of the creation and annihilation operators to the Schwartz inequality $\<\adg a\>_t\big(\<\adg a\>_t+1\big) \geq \<{\adg}^2\>_t\<a^2\>_t$ \cite{Talkner81}.

%%%%%%%%%%%%%%%%%%%%%%%%%%
%       Sectionn         %
%%%%%%%%%%%%%%%%%%%%%%%%%%
\section{Four-dimensional (4D) matrix representation of time evolution operator}
\label{Sec4DU}

We will obtain the solutions to \Eqs{ratemu}-\eqref{ratemunu} indirectly through group theory method. As a preparation, we first obtain the four-dimensional (4D) matrix representation of the time evolution operator. The matrix representation of the generators \eqref{iL0}-\eqref{O+} can be obtained as follows. We introduce the column matrix
%%%
\begin{align} \label{Qv}
    \u{X} &\equiv
        \left(
          \begin{array}{c}
           \d/\d Q \\
           \d/\d r \\
            Q  \\
            r
          \end{array}
        \right)  \,.
\end{align}
%%%
where underlined symbol denotes matrix. Its components are labeled by $X_i$, $i=1, 2, 3, 4$. We find that $[X_i,X_j]=\bt_{ij}$, where $\bt_{ij}$ are the components of the skew-symmetric matrix
%%%
\begin{align}   \label{beta}
    \u{\bt}\equiv \left(\begin{array}{cc}0&\u{I}\\-\u{I}&0\end{array}\right),\qquad
        \u{I}=\left(\begin{array}{cc} 1&0\\0&1 \end{array} \right).
\end{align}
%%%

Under a similarity transformation by the operator
%%%
\begin{align}   \label{SJdef}
    S(J)\equiv\exp(\th J)\,,
\end{align}
%%%
in which $J$ denotes the generic generators in \Eqs{iL0}-\eqref{O+}, $X_i$ transforms as \cite{Simon87}
%%%
\begin{align}   \label{XS}
    X'_i&=S(J)X_iS^{-1}(J)=\sum_j S^{-1}_{ij}(J)X_j\,.
\end{align}	
%%%
$S_{ij}$ are components of the matrix
%%%
\begin{align}   \label{SJmatrix}
    \u{S}(J)\equiv\exp(\th \u{J})\,,
\end{align}	
%%%
whereas $\u{J}$ is the 4D matrix representation of the generator $J$. For infinitesimal transformations, we expand both sides of \Eq{XS} to first order in $\th$ using \Eqs{SJdef} and \eqref{SJmatrix} to obtain
%%%
\begin{align}   \label{SJ}
    [J,X_i]=-\sum_j J_{ij}X_j\,.
\end{align}
%%%
Applying \Eq{SJ} to each of the generators \eqref{iL0}-\eqref{O+}, we can extract the components $J_{ij}$ of the matrix representation of the generators. They are
%%%
\begin{subequations}
\begin{align}   \label{L04d}
    i\u{L}_0 = \frac{i}{2}
        \left(\begin{array}{cc}
            0&\u{\sig}_1\\
            \u{\sig}_1&0
          \end{array}\right),\quad
    i\u{M}_1 = \frac{i}{2}
        \left(\begin{array}{cc}
            0&\u{\sig}_1\\
            -\u{\sig}_1&0
          \end{array}\right), \quad
    i\u{M}_2 = \frac{1}{2}
        \left(\begin{array}{cc}
            -\u{I}&0\\
            0&\u{I}
          \end{array}\right), \quad
    \u{O}_0 = \half\left(\begin{array}{cc}
            -\u{\sig}_3&0\\
            0&\u{\sig}_3
          \end{array}\right),
\end{align}
%%%
%%%
\begin{align}   \label{O+d}
    \u{O}_+ = -\frac{1}{2}
        \left(\begin{array}{cc}
            0&\u{\sig}_d\\
            \u{\sig}_u&0
          \end{array}\right), \quad
     \u{L}_{1+} = \frac{1}{2}
        \left(\begin{array}{cc}
            0&-\u{\sig}_d\\
            \u{\sig}_u&0
          \end{array}\right), \quad
     \u{L}_{2+} = \frac{i}{2}
        \left(\begin{array}{cc}
            -\u{\sig}_-&0\\
            0&\u{\sig}_+
          \end{array}\right),
\end{align}
\end{subequations}
%%%
where $\u{\sig}_i$ denote the matrices
%%%
\begin{subequations}
\begin{align}   \label{sig}
    \u{\sig}_1=\left(\begin{array}{cc} 0&1\\1&0 \end{array} \right),\quad
    \u{\sig}_+=\left(\begin{array}{cc} 0&1\\0&0 \end{array} \right),\quad
    \u{\sig}_-=\left(\begin{array}{cc} 0&0\\1&0 \end{array} \right),\quad
    \u{\sig}_u=\left(\begin{array}{cc} 1&0\\0&0 \end{array} \right),\quad
    \u{\sig}_d=\left(\begin{array}{cc} 0&0\\0&1 \end{array} \right).
\end{align}
\end{subequations}
%%%
The generators \eqref{L04d}-\eqref{O+d} satisfy $\u{\bt}\u{J}=(\u{\bt}\u{J})^\T$, so that the commutation relation $[X'_i,X'_j]=\bt_{ij}$ is preserved. Moreover, they obey the quadratic condition,
%%%
\begin{align}   \label{STbS}
    \u{S}^\T(J)\u{\bt}\u{S}(J)=\u{\bt}\,,
\end{align}
%%%
which means that they belong to the sympletic group in 4D \cite{Gilmore}.

Using the matrix representation of the generators, we show in \ref{SecDecomp} that the time evolution operator can be cast into the following form,
%%%
\begin{subequations}
\begin{align}   \label{expL2D}
    e^{-t\Ku}&=e^{\gam t/2}\bigg[\cosh(\w t/2)\cosh(\gam t/2)\u{I} +\frac{2}{\w\gam}\sinh(\w t/2)\sinh(\gam t/2)\u{H} -\frac{2}{\w}\sinh(\w t/2)\cosh(\gam t/2)\Ku_0 \no
                &\qquad\quad-2\cosh(\w t/2)\sinh(\gam t/2) \u{O}_0
                    -\Sigma_0\u{O}_+-e^{\gam t/2}\Sigma_1\u{L}_{1+}-e^{\gam t/2}\Sigma_2\u{L}_{2+}\bigg]\,,\\
    \Sigv&\equiv \left(\frac{\sinh\big[(\gam-\w)t/2\big]}{\gam-\w}+
            \frac{\sinh\big[(\gam+\w)t/2\big]}{\gam+\w}\right) \ev -\left(\frac{\sinh\big[(\gam-\w)t/2\big]}{\gam-\w}-
            \frac{\sinh\big[(\gam+\w)t/2\big]}{\gam+\w}\right)
            \left(\frac{(\thv\cdot\ev)}{\gam}\thvh+\thvh\wedge\ev\right)\,.\label{Bv}
%    \Thv&\equiv (\thv\cdot\ev)\thv+\gam(\thv\wedge\ev)\,.
%    P_\pm&\equiv \frac{\sinh\big[(\w-\gam)t/2\big]}{\w-\gam}\pm            \frac{\sinh\big[(\w+\gam)t/2\big]}{\w+\gam}\,.
\end{align}
\end{subequations}
%%%
Notice that the prefactor $\exp(\gam t/2)$ in \Eq{expL2D} arises from the $-\gam/2$ term in $\Km_1$ \eqref{K1}.

In \Eq{Bv}, we have introduced three-dimensional Minkowski space vectors with metric signature $(-,+,+)$, labeled by boldfaced letters such as
%%%
\begin{align}   \label{Thv}
    \ev=\eta_0 \e_0 +\eta_1 \e_1 +\eta_2 \e_2\,,
\end{align}
%%%
and similarly for $\thv$ and $\Sigv$. The basis vectors have the scalar products \cite{Simon89}
%%%
\begin{align} \label{e}
    \e_0\cdot\e_0&=-1\,,\quad
    \e_1\cdot\e_1=1\,,\quad
    \e_2\cdot\e_2=1\,,
\end{align}
%%%
and zero otherwise. The cross products are defined in close analogy to the commutation relations of the su(1,1) algebra
%%%
\begin{align}
    \e_0\wedge\e_1&=-\e_2\,,\quad
    \e_1\wedge\e_2=\e_0\,,\quad
    \e_2\wedge\e_0=-\e_1\,,
\end{align}
%%%
and zero otherwise. We note that the cross products differ by an overall sign compared to those introduced in Ref.~\cite{Simon89}. Some useful identities of the vectors are
%%%
\begin{subequations}
\begin{align}   \label{cross}
    \av\cdot(\bv\wedge\cv)&=\av\wedge\bv\cdot\cv=\bv\cdot(\cv\wedge\av)\,,\\
    \av\wedge(\bv\wedge\cv)&=(\av\cdot\bv)\cv-(\av\cdot\cv)\bv\,,\\
    (\av\wedge\bv)^2&=(\av\cdot\bv)^2-\av^2\bv^2\,.
\end{align}
\end{subequations}
%%%

%%%%%%%%%%%%%%%%%%%%%%%%%%
%       Sectionn         %
%%%%%%%%%%%%%%%%%%%%%%%%%%
\section{Baker-Campbell-Hausdorff (BCH) formula of time evolution operator}
\label{SecBCHUt}

In the next step, we propose the following Baker-Campbell-Hausdorff (BCH) formula for the time evolution operator,
%%%
\begin{align}   \label{Kdecompose}
    e^{-t\Km} &=e^{g_2(t)L_{2+}}e^{g_1(t)L_{1+}}e^{g_0(t)O_+}e^{h(t)(O_0-1/2)} e^{m_+(t)M_+}e^{\ln m_0(t)M_0}e^{m_-(t)M_-}\,,
\end{align}
%%%
where it is more convenient to use a different combination of generators for $K_0$, namely,
%%%
\begin{align} \label{Kpm}
    M_0\equiv L_0\,, \qquad M_\pm\equiv M_1\pm iM_2\,,
\end{align}
%%%
because in the 4D matrix representation, we find that $\u{M}_\pm^2=\u{0}$.

Using the Wei-Norman method \cite{Wei63}, we find that the coefficients satisfy the following equations,
%%%
\begin{subequations}
\begin{align}
   \dt{m_-}&=-\half (i\th_1- \th_2) m_0\,,\label{dmmdt}\\
   \dt{m_0}&=-i\th_0 m_0-(i\th_1-\th_2)m_+m_0\,,\label{dm0dt}\\
   \dt{m_+}&=-\half(i\th_1+\th_2)-i\th_0 m_+-\half(i\th_1-\th_2)m_+^2\,,\label{dmpdt}\\
   \dt{h}&=-\gam\,,\label{dhdt}\\
   \dt{\gv}&=-\ev-\gam\gv-\thv\wedge\gv\,.\label{dgdt}
\end{align}
\end{subequations}
%%%
\Eqs{dmmdt}-\eqref{dgdt} still hold if the coefficients in the original generator, $\gam, \th_i, \eta_i,$ have time-dependence. Since we consider only constant coefficients, the generic solution is simpler and can be worked out explicitly.

Instead of solving the set of differential equations directly, we solve for the coefficients by group theory method. We work out the coefficients in the BCH formula in terms of the coefficients $\gam, \th_i, \eta_i$. After writing out the right hand side (RHS) of the BCH formula \eqref{Kdecompose} in the matrix representation, it is equated to the corresponding matrix representation on the RHS of \Eq{expL2D} to yield the desired results.

Noticing that $\u{J}^2=\u{0}$ for the generators except $\u{O}_0$ and $\u{M}_0$, the BCH formula \eqref{Kdecompose} takes the form
%%%
\begin{align}   \label{Kdecomp4D}
    e^{-t\u{K}}&=e^{g_2(t)\u{L}_{2+}}e^{g_1(t)\u{L}_{1+}}e^{g_0(t)\u{O}_+}e^{h(t)(\u{O}_0-1/2)}e^{m_+(t)\u{M}_+}e^{\ln m_0(t)\u{M}_0}e^{m_-(t)\u{M}_-}\no
    &=\left[\u{I}+g_2(t) \u{L}_{2+}\right]\left[\u{I}+g_1(t) \u{L}_{1+}\right]
        \left[\u{I}+g_0(t)\u{O}_+\right] e^{-h(t)/2}\left[\cosh(h(t)/2)\u{I}+2\sinh(h(t)/2) \u{O}_0\right]\no
        &\qquad \times\left[\u{I}+m_+(t)\u{M}_+\right] \left[\cosh\big(\ln\!\!\sqrt{m_0(t)}\big)\u{I}+2\sinh\big(\ln\!\!\sqrt{m_0(t)}\big)\u{M}_0\right]\left[\u{I}+m_-\u{M}_-\right]\,.
\end{align}
%%%
We can then carry out the matrix multiplication straight-forwardly. As shown in \ref{AppMrepresBCH}, after setting the coefficients of the matrices equal to those of \Eq{expL2D}, we obtain the following results,
%%%
\begin{subequations}
\begin{align} \label{ht}
    h(t)&= -\gam t\,,\\
    m_0(t)&= [\cosh(\w t/2)+i\thh_0 \sinh(\w t/2)]^{-2}\,, \label{A0}\\
    m_\pm(t)&=-(i\thh_1\pm\thh_2)\sinh(\w t/2) \sqrt{m_0}\,,\label{coeffKpm}\\
    \gv(t)&= \Sv+\frac{e^{-\gam t}}{\gam}(\thvh\cdot\ev)\thvh
        +\frac{e^{-(\gam-\w)t}}{2(\gam-\w)} \Pv_+(\ev)
        +\frac{e^{-(\gam+\w) t}}{2(\gam+\w)} \Pv_-(\ev)\,,\label{gvt}
\end{align}
\end{subequations}
%%%
with the initial conditions $h(0)=0, m_0(0)=1, m_\pm(0)=0,$ and $\gv(0)=\v{0}$. We have defined a unit vector $\thvh$ with components
%%%
\begin{align}   \label{thh}
    \thh_i\equiv\th_i/\w\,, \qquad \thvh^2=1\,,
\end{align}
%%%
and
%%%
\begin{align}
    %\thvh&\equiv \thv/\w\,,\label{thv}\\
    \label{w}
    \w&\equiv \sqrt{-\th_0^2+\th_1^2+\th_2^2}\,,\\%\sqrt{-\th_0^2+\th_1^2+\th_2^2}\,,\\-\th_0^2+4\th_+ \th_-=\thv^2\,,\\
    \Sv&\equiv \frac{1}{\gam(\gam^2-\w^2)} \left(-\gam^2\ev+(\thv\cdot\ev)\thv +\gam \thv\wedge \ev\right)\,.\label{Sv}
\end{align}
%%%
We note that $\w=4i\w_\text{damp}$. As the coefficients vary, $\Sv$ probes the Minkowski space spanned by the three linearly independent vectors, $\ev, \thv$, and $\thv\wedge\ev$. $\Pv_\pm$ projects $\ev$ into a light-like vector,
%%%
\begin{align}   \label{Pi}
    \Pv_\pm(\ev)&\equiv \ev-\thvh(\thvh\cdot\ev)\mp \thvh\wedge \ev\,,
%    \alh_i&\equiv \frac{\al_i}{\w}\,,
\end{align}
%%%
which lies on a plane Lorentz orthogonal to $\thvh$. The projectors have the following properties,
%%%
\begin{subequations}
\begin{align}   \label{propPi}
    &\Pv_\pm(\ev)\cdot\Pv_\pm(\ev)=0\,,& &(\text{light-like}),\\
    &\thvh\cdot\Pv_\pm(\ev)=0\,,& &(\text{Lorentz orthogonal to $\thvh$}),\\
    &\half\Pv_\pm \left(\half\Pv_\pm(\ev)\right)=\half\Pv_\pm(\ev)\,, & &(\text{idempotent}).
%    &\thvh\wedge\Pv_\pm(\ev)=\mp\Pv_\pm(\ev)\,,\\
%    &\Pv_\pm(\Pv_\mp(\ev))=0\,,\\
%    &\Pv_+(-\ev)=-\Pv_-(\ev)\,.
\end{align}
\end{subequations}
%%%
We can verify straight-forwardly that \Eqs{ht}-\eqref{gvt} are indeed the solutions to \Eqs{dmmdt}-\eqref{dgdt}.

%%%%%%%%%%%%%%%%%%%%%%%%%%
%       Sectionn         %
%%%%%%%%%%%%%%%%%%%%%%%%%%
\section{Solutions of generic master equation}
\label{SecSolMME}

%%%%%%%%%%%%%%%%%%%%%%%%%%
%       Sectionn         %
%%%%%%%%%%%%%%%%%%%%%%%%%%
\subsection{Matrix representation of the density matrices}
\label{SecMRepresrho}

Let us start with an initially normalized density matrix,
%%%
\begin{align}   \label{rho0}
    \rho(Q,r;0)&\equiv\sqrt{\frac{2\mu_0}{\pi}}\rho'(Q,r;0)\,,\qquad
    \rho'(Q,r;0)\equiv e^{-4\mu_0 Q^2/2-\kap_0 iQr-(\mu_0+\nu_0)r^2/2}\,.
\end{align}
%%%
It evolves into
%%%
\begin{align}   \label{timerho}
    \rho(Q,r;t)=e^{-Kt}\rho(Q,r;0)&=\sqrt{\frac{2\mu_0 e^{\gam t}}{\pi}} e^{-K't}\rho'(Q,r;0)\,,
\end{align}
%%%
where we have extracted the factor $\exp(\gam t/2)$ from $\exp(-K t)$ by defining
%%%
\begin{align}   \label{Kp}
    K'\equiv K+\gam/2\,.
\end{align}
%%%
The matrix representation of $\exp(-K' t)$ can be inferred from the first equality of \Eq{Kdecomp4D}. On the other hand, the matrix representation of $\rho'(0)$ can be obtained by using another set of generators
%%%
\begin{align}   \label{genQr}
&\bigg\{\frac{Q^2}{2},\,\, \frac{r^2}{2},\,\, iQr,\,\, Q\frac{\d}{\d Q} +\half,\,\,
 r \frac{\d}{\d r}+\half, \,\, ir\frac{\d}{\d Q},\,\, iQ\frac{\d}{\d r},\,\, \half\frac{\d^2}{\d Q^2},\,\, \half \frac{\d^2}{\d r^2},\,\, i\frac{\d^2}{\d Q \d r}\bigg\}\,,
\end{align}
%%%
whose matrix representations are given in \ref{AppMrepres}. We find that
%%%
\begin{align}   \label{rho4D}
    \u{\rho}'(Q,r;0)&=e^{-4\mu_0 \u{J}(Q^2/2)-\kap_0 \u{J}(iQr)-(\mu_0+\nu_0)\u{J}(r^2/2)}\no
    &=\left[\u{I}-4\mu_0\u{J}\left(\frac{Q^2}{2}\right)\right] \bigg[\u{I}-\kap_0\u{J}\left(iQr\right)\bigg] \left[\u{I}-(\mu_0+\nu_0)\u{J}\left(\frac{r^2}{2}\right)\right]\no
    &=\left(\begin{array}{cc}   \u{I}&-\u{R}_0\\0&\u{I} \end{array}    \right),
\end{align}
%%%
where
%%%
\begin{align}   \label{rhoR}
    \u{R}_0\equiv\left(\begin{array}{cc}4\mu_0&i\kap_0\\i\kap_0&\mu_0+\nu_0\end{array}\right).
\end{align}
%%%
We can then multiply the matrices to obtain
%%%
\begin{align}   \label{Urho4D}
    e^{-\u{K}'t}\u{\rho}'(Q,r;0)&\equiv\left(\begin{array}{cc}   \u{D}_{11}&\u{D}_{12}\\\u{D}_{21}&\u{D}_{22} \end{array}    \right),
\end{align}
%%%
where $\u{D}_{ij}$ are matrices that are functions of $\mu_0, \nu_0, \kap_0$ and the coefficients from the BCH formula \eqref{Kdecomp4D}. The $\u{D}_{ij}$ can be worked out explicitly.

Since $e^{-\u{K}'t}\u{\rho}'(0)$ belongs to the complex symplectic group \cite{Tay06}, it satisfies the quadratic condition \eqref{STbS} \cite{Gilmore,Simon87}, which gives rise to the following matrix relations \cite{Simon87},
%%%
\begin{subequations}
\begin{align}   \label{ABCD}
    \u{D}_{11}^\T \u{D}_{21}=(\u{D}_{11}^\T\u{D}_{21})^\T\,,\\
    \u{D}_{12}^\T \u{D}_{22}=(\u{D}_{12}^\T\u{D}_{22})^\T\,,\\
    \u{D}_{11}^\T \u{D}_{22}-\u{D}_{21}^\T\u{D}_{12}=\u{I}\,,\label{ABCDc}
\end{align}
\end{subequations}
%%%
where the superscript $\T$ denotes matrix transposition.

%%%%%%%%%%%%%%%%%%%%%%%%%%
%       Sectionn         %
%%%%%%%%%%%%%%%%%%%%%%%%%%
\subsection{Baker-Campbell-Hausdorff (BCH) formula of density matrix}
\label{SecDecompUrho}

As the final step before we arrive at the solution, we propose another form of BCH formula for $\exp(-K't)\rho'(0)$,
%%%
\begin{align}   \label{Urhodecomp}
    e^{-K't}\rho'(0)
    &\equiv e^{-4\mu(t)Q^2/2-\kap(t)iQr-[\mu(t)+\nu(t)]r^2/2}
            e^{u(t)\left(Q{\d}/{\d Q}+\half\right)}
            e^{v(t)\left(r{\d}/{\d r}+\half\right)}
            e^{p(t)ir{\d}/{\d Q}}
            e^{q(t)iQ{\d}/{\d r}}
            e^{{f_1(t)}/{2}{\d^2}/{\d Q^2}}
            e^{if_3(t){\d^2}/{\d Q \d r}}
            e^{{f_2(t)}/{2}{\d^2}/{\d r^2}}\,.
\end{align}
%%%
The matrix representation of the RHS of \Eq{Urhodecomp} are respectively given by
%%%
\begin{subequations}
\begin{align}   \label{4Drhoa}
    e^{-4\mu(t)\u{J}(Q^2/2)-\kap(t)\u{J}(iQr)-[\mu(t)+\nu(t)]\u{J}(r^2/2)}&=\left(\begin{array}{cc}   \u{I}&-\u{R}\\0&\u{I} \end{array}    \right),\qquad
            \u{R}\equiv \left(\begin{array}{cc}   4\mu(t)&i\kap(t)\\i\kap(t)&\mu(t)+\nu(t) \end{array}    \right),
\end{align}
%%%
%%%
\begin{align}   \label{Vmatrix}
    e^{u(t)\u{J}\left(Q{\d}/{\d Q}+\half\right)}
            e^{v(t)\u{J}\left(r{\d}/{\d r}+\half\right)}&=\left(\begin{array}{cc}   \u{W}&0\\0&\u{W}^{-1} \end{array}    \right),\qquad
            \u{W}\equiv \left(\begin{array}{cc}   e^{u(t)}&0\\0&e^{v(t)} \end{array}    \right),
\end{align}
%%%
%%%
\begin{align}   \label{Xmatrix}
    e^{p(t)\u{J}(ir{\d}/{\d Q})}e^{q(t)\u{J}(iQ{\d}/{\d r})}&=\left(\begin{array}{cc}   \u{X}&0\\0&\left(\u{X}^\T\right)^{-1} \end{array}    \right),\qquad
            \u{X}\equiv \left(\begin{array}{cc}   1&p(t)\\q(t)&1+p(t)q(t) \end{array}    \right),
\end{align}
%%%
%%%
\begin{align}   \label{Fmatrix}
     e^{{f_1(t)}/{2}\u{J}(\d^2/\d Q^2)}
            e^{if_3(t)\u{J}(\d^2/\d Q \d r)}
            e^{{f_2(t)}/{2}\u{J}(\d^2/\d r^2)}&=\left(\begin{array}{cc}   \u{I}&0\\-\u{F}&\u{I}	 \end{array}    \right),\qquad
            \u{F}\equiv \left(\begin{array}{cc}   f_1(t)&if_3(t)\\if_3(t)&f_2(t) \end{array}    \right).
\end{align}
\end{subequations}
%%%
After carrying out the matrix multiplications on the RHS of \Eqs{Urhodecomp} using \Eqs{4Drhoa}-\eqref{Fmatrix}, we equate the resulting matrix to \Eq{Urho4D}. This procedure yields four matrix equations,
%%%
\begin{subequations}
\begin{align}   \label{4Meq}
    \u{D}_{11}&=\u{W}\u{X}+\u{R}(\u{X}\u{W})^{-1}\u{F}\,,\\
    \u{D}_{12}&=-\u{R}(\u{X}\u{W})^{-1}\,,\\
    \u{D}_{21}&=-(\u{X}\u{W})^{-1}\u{F}\,,\\
    \u{D}_{22}&=(\u{X}\u{W})^{-1}\,.
\end{align}
\end{subequations}
%%%
The unknown matrices on the RHS can be solved in terms of the $\u{D}_{ij}$ matrices,
%%%
\begin{subequations}
\begin{align}   \label{RprimeUXF}
    \u{R}&=-\u{D}_{12}\u{D}_{22}^{-1}\,,\\
    \u{F}&=-\u{D}_{22}^{-1}\u{D}_{21}\,,\\
    \u{X}\u{W}&=\u{D}_{22}^{-1}\,,\label{UX}
\end{align}
\end{subequations}
%%%
which give the ten coefficients in the BCH formula \eqref{Urhodecomp}. \Eq{4Meq} reproduces \Eq{ABCDc}, so it does not provide new information. Since the identity \eqref{Urhodecomp} is obtained in the defining 4D matrix representation, it is also valid in all representations \cite{Gilmore}, including the infinite dimensional representation we start with.

Lastly, we operate \Eq{Urhodecomp} on constant so that the exponentials containing differential operators on the RHS of \Eq{Urhodecomp} reduce to identity. Substituting the results back into \Eq{timerho}, we are left with a simple expression,
%%%
\begin{align}   \label{rhoprimeUrho}
    \rho(Q,r;t)&=\sqrt{\frac{2\mu_0e^{\gam t}e^{u(t)+v(t)}}{\pi}}
            e^{-4\mu(t){Q^2}/{2}-\kap(t)iQr-[\mu(t)+\nu(t)]{r^2}/{2}}\,,
\end{align}
%%%
where $\mu(t), \kap(t)$ and $\nu(t)$ can be worked out directly from \Eq{RprimeUXF} by matrix multiplication, whereas $\exp[u(t)+v(t)]$ can be solved explicitly from \Eq{UX} using \Eqs{Vmatrix} and \eqref{Xmatrix} to yield $\exp[u(t)+v(t)]=1/\text{det}\u{D}_{22}$. By inspecting the first equation in \Eq{evomu}, \Eqs{Dt} and \eqref{mu'}, we then realize that the factor $\mu_0e^{\gam t}e^{u(t)+v(t)}$ under the square root is none other than $\mu(t)$. In this way we recover $\rho(Q,r;t)$ \eqref{rhoGaussian}, which shows the consistency of this method. We note that a comparison of the RHS of \Eqs{Urhodecomp} and \eqref{rhoprimeUrho} reveals the fact that the density matrices has an infinite degeneracy in the group space.

%%%%%%%%%%%%%%%%%%%%%%%%%%
%       Sectionn         %
%%%%%%%%%%%%%%%%%%%%%%%%%%
\subsection{Generic solution}
\label{SecGensol}

The generic solution can be summarised as
%%%
%\begin{subequations}
\begin{align}   \label{evomu}
   \mu(t)\equiv\frac{\mu'(t)}{D(t)}\,, \qquad
   \kap(t)\equiv\frac{\kap'(t)}{D(t)}\,,\qquad
   \nu(t)\equiv\frac{\nu'(t)}{D(t)}\,,%\label{evomunu}
\end{align}
%\end{subequations}
%%%
with the denominator
%%%
\begin{subequations}
\begin{align}   \label{Dt}
   D(t)&\equiv \text{det}\u{D}_{22}=2\mu_0e^{\gam t}\big[\g_0(t)-\g_1(t)\big] -C'+\frac{ e^{\w t}}{2}(1+C'- B')+\frac{ e^{-\w t}}{2}(1+C'+B')\,,\\
%   A'&\equiv 1-C'\,,\\
   B'&\equiv \thh_2+\kap_0(\thh_0-\thh_1)\,,\\
   C'&\equiv (\thh_0-\thh_1)\Phi(\thvh)\,,\\
   \Phi(\thvh)&\equiv \half(\thh_0+\thh_1)+\half\Del_0^2(\thh_0-\thh_1)+\kap_0 \thh_2\,,\\
   \Del_0^2&\equiv \text{det}\u{R}_0=4\mu_0(\mu_0+\nu_0)+\kap_0^2\,.
\end{align}
\end{subequations}
%%%
The numerators are
%%%
\begin{subequations}
\begin{align}   \label{mu'}
   \mu'(t) &= \mu_0 e^{\gam t}\,,\\
   \kap'(t) &= -2\mu_0 e^{\gam t} \g_2(t)  + C +\frac{ e^{\w t}}{2}(\kap_0-C-B)
                +\frac{ e^{-\w t}}{2}(\kap_0-C+B)\,,\\
   \nu'(t)&= (\mu_0+\nu_0)e^{-\gam t}-\mu_0 e^{\gam t}\big[\gv^2(t)+1\big]
        + [\thvh\cdot\gv(t)] \Phi(\thvh)+\frac{ e^{\w t}}{2} \Phi\big(\Pv_-[\gv(t)]\big)
            +\frac{ e^{-\w t}}{2} \Phi\big(\Pv_+[\gv(t)]\big)\,,\\
%   A&\equiv \kap_0+C\,,\\
   B&\equiv -\half(\thh_0+\thh_1)+\half\Del_0^2(\thh_0-\thh_1)\,,\\
   C&\equiv \thh_2 \Phi(\thvh)\,.\label{C}
\end{align}
\end{subequations}
%%%
In \ref{AppEqMotProof}, we discuss how we verify that \Eqs{evomu} with \Eqs{Dt}-\eqref{C} indeed solve the nonlinear equations \eqref{ratemu}-\eqref{ratemunu}.

%%%%%%%%%%%%%%%%%%%%%%%%%%
%       Sectionn         %
%%%%%%%%%%%%%%%%%%%%%%%%%%
\section{Properties of stationary states}
\label{SecBehavSol}

Although the exact solution enables us to follow the time evolution of the density matrices, the expressions are very complicated to analyze. We instead discuss the effects of the coefficients on the stationary states by requiring their existence and positive semidefiniteness. We will also require the stationary states to satisfy a factorized condition in their coordinates. The latter requirement is strong enough to produce the known master equations and their generalizations.

%%%%%%%%%%%%%%%%%%%%%%%%%%
%       Sectionn         %
%%%%%%%%%%%%%%%%%%%%%%%%%%
\subsection{Existence of stationary states}
\label{SecStabSt}

The existence of the stationary states depends on the sign of the damping constant $\gam$. When $\gam$ takes on negative value, from \Eqs{gvt}, \eqref{evomu}, \eqref{Dt}, and \eqref{mu'}, we deduce that in the $t\rightarrow \infty$ limit, the dominant terms that govern the following expressions are
%%%
\begin{subequations}
\begin{align}   \label{Dinf}
    e^{\gam t}\gv(t)&\rightarrow \frac{\thvh\cdot\ev}{\gam}\thvh +\frac{e^{\w t}}{2(\gam-\w)} \Pv_+(\ev) +\frac{e^{-\w t}}{2(\gam+\w)} \Pv_-(\ev) \,,\\
    D(t)&\rightarrow  e^{\gam t}\big[g_0(t)-g_1(t)\big] -C'+\half e^{\w t}(1+C'- B') +\half e^{-\w t}(1+C'+ B')\,.
\end{align}
%%%
Hence, in this limit $\mu(t)$ is proportional to
%%%
\begin{align}
    \mu(t)&=\frac{\mu_0 e^{\gam t}}{D(t)}\propto e^{-|\gam|t-\w t}
\end{align}
\end{subequations}
%%%
regardless of real or imaginary $\w$. As a result, the probability distribution function of the oscillator \eqref{PDF} vanishes in the limit $t\rightarrow \infty$ and the solution does not exist.

For positive $\gam$, we recall from the discussion of Section \ref{SecTmEvo} that both the underdamped and critically damped oscillators have imaginary $\w$. Therefore, in the limit $t\rightarrow \infty$, we obtain the following behaviors,
%%%
\begin{align}   \label{gvinfimag}
    e^{\gam t}\gv(t)&\rightarrow e^{\gam t}\Sv\,,\qquad
    D(t)\rightarrow 2\mu_0 e^{\gam t}(\Gam_0-\Gam_1)\,.
\end{align}
%%%
As a result, \Eqs{evomu}-\eqref{C} give
%%%
\begin{subequations}
\begin{align}   \label{muinf2}
    \mu_\text{st}&=\frac{1}{2(\Gam_0-\Gam_1)}\,,\\
    \nu_\text{st}&=\frac{-\Sv^2-1}{2(\Gam_0-\Gam_1)}\,,\label{munuinf2}\\
    \kap_\text{st}&=\frac{-\Gam_2}{\Gam_0-\Gam_1}\,,\label{kinf2}
\end{align}
\end{subequations}
%%%
for the stationary states.

On the other hand, for an overdamped oscillator with real $\w$, only $\w<\gam$ produces meaningful stationary states. In this case, a similar analysis shows that it behaves like \Eq{gvinfimag}, and hence its solution exhibits similar behaviours to the underdamped and critically damped oscillator \eqref{muinf2}-\eqref{kinf2}. When an overdamped oscillator has $\w>\gam$ or $\w=\gam$, its $\mu(t)$ behaves like $e^{-(\w-\gam)t}$ and $e^{-\gam t}$ in the limit $t \rightarrow \infty$, respectively. Hence the solution does not exist.

In summary, the stationary states of the density matrices of the underdamped, critically damped, and overdamped oscillator have similar behaviours. Stable solutions exist only when $\gam>0$, and in the case of the overdamped oscillator, we need to limit the frequency to $\w<\gam$ for stable solutions to exist.

%%%%%%%%%%%%%%%%%%%%%%%%%%
%       Sectionn         %
%%%%%%%%%%%%%%%%%%%%%%%%%%
\subsection{Positive semidefinite stationary states}
\label{SecPos}

The coefficient $\mu(t)$ characterizes the probability distribution function \eqref{PDF}. A well-behaved probability distribution function requires $\mu(t)>0$ \eqref{nscond}, or equivalently,
%%%
\begin{align}   \label{hermcond}
    \Gam_0-\Gam_1 > 0\,.
\end{align}
%%%
As discussed in the previous section, for the overdamped oscillator we need to consider only $\w<\gam$. Consequently, in all types of oscillator, the prefactor on the RHS of \Eq{Sv} for $\Sv$ can be written as a positive quantity, $1/\big[\gam(\gam^2-\thv^2)\big]$, for real or imaginary $\w$, multiplied by $-\gam^2 \ev+(\thv\cdot\ev)\thv+\gam \thv\wedge\ev$. After rewriting this expression in component form, \Eq{hermcond} is equivalent to
%%%
\begin{align}   \label{G0-G1}
    \gam(-\eta_0+\eta_1)(\gam-\th_2)+(\th_0-\th_1)(-\th_0\eta_0+\th_1\eta_1)-\eta_2(\th_0-\th_1)(\gam-\th_2)>0\,,
\end{align}
%%%
for both real or imaginary $\w$.

The sufficient condition for the positive semidefiniteness of the stationary states is provided by $\nu(t)\geq 0$ \eqref{nscond}, or from \Eq{munuinf2},
%%%
\begin{align} \label{poscond}
    -\Sv^2 \geq 1\,.
\end{align}
%%%
Similar to the analysis of $\Sv$, $-\Sv^2$ can be written as a positive quantity, $1/\big[\gam^2(\gam^2-\thv^2)\big]$, for real or imaginary $\w$, multiplied by $(\thv\cdot\ev)^2-\gam^2\ev^2$. This means that we require
%%%
\begin{align}   \label{Svsqasymp}
    (\thv\cdot\ev)^2-\gam^2\ev^2= (-\th_0\eta_0+\th_1\eta_1+\th_2\eta_2)^2+\gam^2(\eta_0^2-\eta_1^2-\eta_2^2)\geq \gam^2(\gam^2-\thv^2)\geq 0\,.
\end{align}
%%%
Although the $\th_i$ form the unitary part of the reduced dynamics, they affect the positivity of the stationary states through the $\thv\cdot\ev$ term.

%%%%%%%%%%%%%%%%%%%%%%%%%%
%       Sectionn         %
%%%%%%%%%%%%%%%%%%%%%%%%%%
\subsection{Factorized condition}
\label{SecSep}

The Wigner functions \cite{Gardiner} of the Gaussian density matrix \eqref{rhoGaussian} takes the form,
%%%
\begin{align}   \label{Wigner}
    W(Q, P; t)&= \frac{1}{\pi}\sqrt{\frac{\mu(t)}{\mu(t)+\nu(t)}} \exp\left({-\frac{4\mu(t)[\mu(t)+\nu(t)]+\kap^2(t)}{2[\mu(t)+\nu(t)]}Q^2 -\frac{\kap(t)}{\mu(t)+\nu(t)}QP -\frac{1}{2[\mu(t)+\nu(t)]}P^2}\right)\,.
\end{align}
%%%
The coefficient $\kap(t)$ determines whether or not the Wigner function can be factorized into two separate functions of $Q$ and $P$. If $\kap(t)$ vanishes, we have
%%%
\begin{align}   \label{WSep}
    W(Q, P; t)&\equiv f(Q)g(P)\,,\\
    f(Q)&\equiv\sqrt{\frac{\mu(t)}{\pi}} e^{-2\mu(t)Q^2}\,,\\
    g(P)&\equiv \frac{e^{-P^2/2[\mu(t)+\nu(t)]}}{\sqrt{\pi[\mu(t)+\nu(t)]}} \,.
\end{align}
%%%
We note that $g(P)$ is the momentum space distribution function conjugates to the position space distribution function $f(Q)$ only for pure states. If stationary states satisfy
%%%
\begin{align}   \label{G2=0}
    \Gam_2&=0\,,
\end{align}
%%%
then their density matrices are factorized in the $Q, r$ coordinates, so do their Wigner function \eqref{WSep}. It is in this sense that we call \Eq{G2=0} a factorized condition. In terms of the coefficients of the master equation, \Eq{G2=0} can be written as
%%%
\begin{align}   \label{SepCond}
    \eta_2(\th_2^2-\gam^2)+\eta_0(-\th_0\th_2+\gam\th_1) +\eta_1(\th_1\th_2-\gam\th_0)=0\,.
\end{align}
%%%

The Gibbs states,
%%%
\begin{align}   \label{Gibbs}
    \rho_\text{Gibbs}(Q,r)=\frac{1}{\sqrt{2\pi b}}e^{-Q^2/2b-b r^2/2}\,,
\end{align}
%%%
which are the stationary states of a system in thermal equilibrium with a thermal reservoir, are examples of stationary states obeying the factorized condition. This can be seen by comparing \Eqs{muinf2}-\eqref{kinf2} with $\rho_\text{Gibbs}$ to yield
%%%
\begin{align}   \label{GibbsGam}
     \text{Gibbs state:} \qquad \Gam_0=2b, \qquad \Gam_1=\Gam_2=0\,.
\end{align}
%%%
Examples of master equations with factorized stationary states are the Kossakowski-Lindblad (KL) equation \cite{Kossa76,Lindblad76}, and the Caldeira-Leggett (CL) equation \cite{CL83}, with generators given by
%%%
\begin{align}   \label{KKL}
    K_\KL(Q, r) &= i\w_0\left(-\frac{\d^2}{\d Q\d r}+Qr\right)-\frac{\gam}{2}\bigg(\frac{\d}{\d Q}Q -r\frac{\d}{\d r}\bigg) -b\frac{\gam}{2}\left(\frac{\d^2}{\d Q^2}-r^2\right),  \\
    K_\CL(Q, r) &= i\w_0\left(-\frac{\d^2}{\d Q\d r}+Qr\right) +i\frac{\th_1}{2}\left(\frac{\d^2}{\d Q\d r}+Qr\right)
                +\gam r\frac{\d}{\d r} + b_\text{CL} \gam r^2\,,\label{KCL}
\end{align}
%%%
respectively. The KL equation is found in quantum optical systems, whereas the CL equation is used to study quantum Brownian motion. We find that
%%%
\begin{align}   \label{rhoequi}
    &\text{KL:} &\qquad &\Gam_0=2b\,, &\qquad &\Gam_1=0\,, &\qquad &\Gam_2=0\,,\\
    &\text{CL:} &\qquad &\Gam_0=2b\frac{2\w_0 }{2\w_0+\th_1}\,, &\qquad &\Gam_1=2b\frac{\th_1}{2\w_0+\th_1}\,, &\qquad &\Gam_2=0\,,
\end{align}
%%%
where
%%%
\begin{align}   \label{b}
    b&=%\frac{1}{2} \coth \frac{\w_0}{2kT}
    \half +\frac{1}{e^{\w_0/2kT}-1}\geq 1/2\,,\\
    b_\CL&=kT/\w_0\,,  \label{bCL}
\end{align}
%%%
in which $b_\text{CL}$ is the high temperature limit of $b$, and we use the units $\hbar=1$. Moreover, for $\th_1=0$, the stationary states of the CL equation are the Gibbs states.

When the system is in contact with non-standard reservoir, or when the system is strongly correlated to the reservoir, the stationary states might not be the Gibbs states. For instance, the Hu-Paz-Zhang (HPZ) equation \cite{HPZ92} with the generator
%%%
\begin{align}   \label{KHPZ}
    K_\HPZ(Q, r) &= i\w_0\left(-\frac{\d^2}{\d Q\d r}+Qr\right) +i\frac{\th_1}{2}\left(\frac{\d^2}{\d Q\d r}+Qr\right)
            +\gam r\frac{\d}{\d r} +b \gam  r^2 + i dr\frac{\d}{\d Q}\,,
\end{align}
%%%
is obtained under quite general conditions, including strong coupling to the reservoir. Though its stationary states are not the Gibbs states, they satisfy the factorized condition,
%%%
\begin{align}
    \text{HPZ:} \qquad
        \Gam_0=2b\frac{2\w_0-\eta_2/2b}{2\w_0+\th_1}\,, &\qquad \Gam_1=2b\frac{\th_1+\eta_2/2b}{2\w_0+\th_1}\,, \qquad \Gam_2=0\,.
\end{align}
%%%

In the following discussion, we are going to impose the factorized condition on the generic stationary state. We find that it provides a systematic way to classify the master equations.

%%%%%%%%%%%%%%%%%%%%%%%%%%
%       Sectionn         %
%%%%%%%%%%%%%%%%%%%%%%%%%%
\subsection{General requirements on the coefficients}
\label{SecGenReq}

Before imposing the factorized condition on the stationary states, we first discuss some general requirements on the coefficients of the master equations. As discussed in Section \ref{SecTmEvo}, we observe that $\th_0>0$ as it is related to the natural frequency of the oscillator. We assume that the coefficient $\th_1$ renormalizes or shifts the natural frequency in a small amount, so that $|\th_1|<|\th_0|$, or $\th_0-\th_1>0$.

From the discussion in Section \ref{SecTmEvo}, the $\d^2/\d Q^2$-term causes diffusion-like effect on the probability distribution functions. Its coefficient $-\eta_0+\eta_1$ is required to be positive semidefinite,
%%%
\begin{align}   \label{dQ>=0}
    -\eta_0+\eta_1\geq 0\,.
\end{align}
%%%
On the other hand, the $r^2$-term causes decoherence \cite{Zurek91}. When this term becomes dominant at large $r$, it introduces to the density matrices a factor $\exp[(\eta_0+\eta_1)r^2 t]$ that decoheres their off-diagonal components. We then require
%%%
\begin{align}   \label{r2<=0}
    \eta_0+\eta_1\leq 0\,.
\end{align}
%%%
From \Eqs{dQ>=0} and \eqref{r2<=0} we conclude that $\eta_0$ and $\eta_1$ satisfy the condition,
%%%
\begin{align}   \label{diff+decoh}
    \eta_0\leq 0\,, \qquad |\eta_1|\leq|\eta_0|\,.
\end{align}
%%%
Due to \Eq{diff+decoh} and the assumption that $|\th_1|<|\th_0|$, the effect of $\eta_1$ is usually less prominent than $\eta_0$.

%%%%%%%%%%%%%%%%%%%%%%%%%%
%       Sectionn         %
%%%%%%%%%%%%%%%%%%%%%%%%%%
\subsection{Comparison with positive conditions obtained from the uncertainty principle}
\label{SecUncert}

In Section \ref{Sec2ndmoment} we found that by requiring second moments to satisfy the uncertainty principle, we arrive at the same necessary and sufficient condition \eqref{nscond} for the positive semidefiniteness of $\rho(t)$. On the other hand, Ref.~\cite{Dekker84} considered the time evolution of damped oscillator under the generator (in our notations)
%%%
\begin{align}   \label{DekkerK}
    K_\text{DV}=i\w_0\left(-\frac{\d^2}{\d Q\d r}+Qr\right) +2\lambda r\frac{\d}{\d r} -D_{qq}\frac{\d^2}{\d Q^2} +D_{pp} r^2 +i(D_{qp}+D_{pq})r\frac{\d}{\d Q}\,,
\end{align}
%%%
where $D_{pq}=D_{qp}$ and for simplicity, we assume that that there is no frequency shift to the oscillator. It was deduced that the uncertainty principle then imposes the following constraints on the diffusion coefficients \cite{Dekker84}
%%%
\begin{align}   \label{diffconstraint}
    D_{qq}>0\,, \qquad D_{pp}>0\,, \qquad D_{qq}D_{pp}-D_{qp}D_{pq}\geq \frac{\lam^2}{4}\,.
\end{align}
%%%
Consequently, it was concluded that the CL and HPZ equation must be rejected because they do not satisfy \Eq{diffconstraint}.

Comparing \Eq{DekkerK} with \Eqs{K}-\eqref{O+}, we find that
%%%
\begin{subequations}
\begin{align}   \label{Dekkercoeff}
    \th_0&=2\w_0\,,\qquad \th_1=0\,,\qquad \th_2=-\gam\,,\qquad \lam=\frac{\gam}{2}\,,\\
    D_{qq}&=-\frac{1}{4}(\eta_0-\eta_1)\,,\qquad
    D_{pp}=-\frac{1}{4}(\eta_0+\eta_1)\,,\qquad
    D_{qp}=-\frac{1}{4}\eta_2\,.
\end{align}
\end{subequations}
%%%
From \Eq{Dekkercoeff}, $\th_2=-\gam$ shows that the master equation considered is closely related to the class of CL and HPZ equations, cf. Table 1 at the end of Section \ref{SecMEsep}.

The constraints on $D_{qq}$ and $D_{pp}$ in \Eq{diffconstraint} are much stronger than \Eqs{dQ>=0} and \eqref{r2<=0}. The KL, HPZ equation and their conjugates could allow equalities in these conditions, cf. Table 1. We will also find that the equalities in \Eqs{dQ>=0} and \eqref{r2<=0} also appear in other master equations that satisfy the factorized condition. Moreover, from the relation $16(D_{qq}D_{pp}-D_{qp}D_{pq})=\eta_0^2-\eta_1^2-\eta_2^2$, the third constraint in \Eq{diffconstraint} can be written as
%%%
\begin{align}   \label{Dqp}
    -\ev^2\geq \gam^2\,,\qquad \text{(positive condition obtained in Ref.~[25]).}
\end{align}
%%%

On the other hand, in Section \ref{Sec2ndmoment} we have shown that by requiring the second moments to satisfy the uncertainty principle at all time, we recover the positive condition \eqref{nscond}. When this condition is applied to the stationary state of generic master equation, we obtain condition \eqref{Svsqasymp} which can also be written as
%%%
\begin{align}   \label{pos}
    \frac{1}{\gam^2}(\thv\cdot\ev)^2+\thv^2-\ev^2\geq \gam^2\,,\qquad \text{(positive condition for generic stationary states).}
\end{align}
%%%
\Eq{pos} reveals that on the left hand side of the inequality, \Eq{pos} contains two extra terms involving coefficients from the unitary part of the reduced dynamics compared to \Eq{Dqp}. Though the CL and HPZ equation do not satisfy \Eq{Dqp}, there are coefficients of the CL and HPZ equation that could satisfy \Eq{pos}. Furthermore, in the numerical analysis in Section \ref{SecMEsep}, we show that there exist initial state and coefficients of the CL and HPZ equation that could generate positive evolution for all time.

The discrepancy between \Eqs{Dqp} and \eqref{pos} arises because the third inequality of \Eq{diffconstraint} or \eqref{Dqp} is too restrictive. Even if $0\leq D_{qq}D_{pp}-D_{qp}D_{pq}< \lam^2/4$, there are coefficients of the master equation for which the sum of the first two positive terms in Eq.~(10) of Ref.~\cite{Dekker84} are larger than its third negative term, to ensure that Eq.~(10) of Ref.~\cite{Dekker84} is satisfied. Therefore, the positive semidefinite conditions we obtain are more general.

%%%%%%%%%%%%%%%%%%%%%%%%%%
%       Sectionn         %
%%%%%%%%%%%%%%%%%%%%%%%%%%
\section{Master equations with factorized stationary states}
\label{SecMEsep}

%%%%%%%%%%%%%%%%%%%%%%%%%%
%       Sectionn         %
%%%%%%%%%%%%%%%%%%%%%%%%%%
%\subsection{Imposing the factorized condition}
%\label{SecMEFac}

Let us now impose the factorized condition \eqref{G2=0} on generic stationary state. Keeping in mind that $\eta_i$ are independent coefficients, the master equations can be divided into two classes, according to either $\eta_2\neq0$ or $\eta_2=0$.

%%%%%%%%%%%%%%%%%%%%%%%%%%
%       Sectionn         %
%%%%%%%%%%%%%%%%%%%%%%%%%%
%\subsection{Case I: $\eta_2\neq0$.}
%\label{SecI}

\begin{enumerate}[{(I)}]
\item $\eta_2\neq0$.

We require the coefficient of $\eta_2$ in \Eq{SepCond} to vanish, which divides the master equations into two classes, depending on whether $\th_2=-\gam$ or $\gam$. In both cases, a zero $\eta_0$ of the master equation is not consistent the assumption that the frequency shift is small compared to the natural frequency. For, if $\eta_0=0$, \Eq{diff+decoh} requires $\eta_1=0$. As a result, the positive condition $-\Sv^2\geq1$ or \Eq{Svsqasymp} requires $\th_1=\pm\th_0$, which contradicts the assumption. As a result, we consider only $\eta_0\neq0$.

\begin{enumerate}[(A)]
\item $\th_2=-\gam$. HPZ equation.

The factorized condition further gives
%%%
\begin{align}
    \eta_1=\eta_0\,.
\end{align}
%%%
The equation does not contain the diffusion-like $\d^2/\d Q^2$-term. It produces the HPZ equation \eqref{KHPZ}. In the Wigner representation, it is
%%%
\begin{align}   \label{HPZQP}
    K_\text{HPZ}(Q,P)=\frac{\th_0}{2}\left(P\frac{\d}{\d Q}-Q\frac{\d}{\d P}\right) -\frac{\th_1}{2}\left(P\frac{\d}{\d Q}+Q\frac{\d}{\d P}\right) -\gam \frac{\d}{\d P}P +\frac{\eta_0}{2}\frac{\d^2}{\d P^2}+\frac{\eta_2}{2}\frac{\d^2}{\d Q \d P}\,.
%    K_\text{HPZ}=-\frac{1}{2}(\th_0-\th_1)i\frac{\d^2}{\d Q\d r}+\frac{1}{2}(\th_0+\th_1)iQr+\gam r\frac{\d}{\d r}-\frac{\eta_0}{2}r^2-\frac{\eta_2}{2}ir\frac{\d}{\d Q}\,.
\end{align}
%%%
Furthermore, the conditions \eqref{G0-G1} and \eqref{Svsqasymp} give the following constraints on the coefficients of stationary states that are positive semidefinite,
%%%
\begin{align}   \label{IBG0G1}
    \Gam_0-\Gam_1 > 0: \qquad &\eta_2<(\th_0-\th_1)\frac{|\eta_0|}{2\gam}\,,\\
    -\Sv^2\geq 1: \qquad & \eta_2\leq (\th_0-\th_1) \frac{|\eta_0|}{2\gam}-(\th_0+\th_1)\frac{\gam}{2|\eta_0|} \,.\label{IBG2}
\end{align}
%%%
The requirement on the decoherence $r^2$ term, or the $-\d^2/\d P^2$ term in the Wigner representation, $\eta_0+\eta_1\leq 0$, then gives $\eta_0\leq 0$. Since $\eta_0=0$ is not permitted as discussed at the end of Section \ref{SecGenReq}, $\eta_0<0$ for the HPZ equation. The HPZ equation with $\eta_2=0$ coincides with the CL equation \eqref{CLQP}.

\item $\th_2=\gam$. Conjugate of HPZ equation.

The factorized condition \eqref{SepCond} reduces to $(-\th_0+\th_1)(\eta_0+\eta_1)=0$. The independence of $\th_0, \th_1$ then requires
%%%
\begin{align}
    \eta_1=-\eta_0\,,
\end{align}
%%%
which means that the equation does not contain the $r^2$ decoherence term. In the Wigner representation, this class of equation has the generator
%%%
\begin{align}   \label{diffHPZ}
    K_\text{cHPZ}(Q,P)=\frac{\th_0}{2}\left(P\frac{\d}{\d Q}-Q\frac{\d}{\d P}\right) -\frac{\th_1}{2}\left(P\frac{\d}{\d Q}+Q\frac{\d}{\d P}\right) -\gam \frac{\d}{\d Q}Q+\frac{\eta_0}{2}\frac{\d^2}{\d Q^2}+\frac{\eta_2}{2}\frac{\d^2}{\d Q \d P}\,.
%    K_\text{dHPZ}=-\frac{1}{2}(\th_0-\th_1)i\frac{\d^2}{\d Q\d r}+\frac{1}{2}(\th_0+\th_1)iQr-\gam \frac{\d}{\d Q}Q+\frac{\eta_0}{2}\frac{\d^2}{\d Q^2}-\frac{\eta_2}{2}ir\frac{\d}{\d Q}\,.
\end{align}
%%%
We shall call $K_\text{cHPZ}$ the conjugate of the HPZ equation. The positive semidefinite conditions \eqref{G0-G1} and \eqref{Svsqasymp} then give
%%%
\begin{align}   \label{IAG0G1}
    \Gam_0-\Gam_1 > 0: \qquad &\eta_0<0\,,\\
    -\Sv^2\geq 1: \qquad & -(\th_0+\th_1) \frac{|\eta_0|}{2\gam}+(\th_0-\th_1)\frac{\gam}{2|\eta_0|}\leq \eta_2\,.\label{IAG2}
\end{align}
%%%
The conjugate equation also satisfies the requirements on the diffusion coefficients, \Eqs{dQ>=0}-\eqref{diff+decoh}.

\end{enumerate}
\end{enumerate}

%%%
\begin{figure}[t]
  \centering
\includegraphics[width=3.2in]{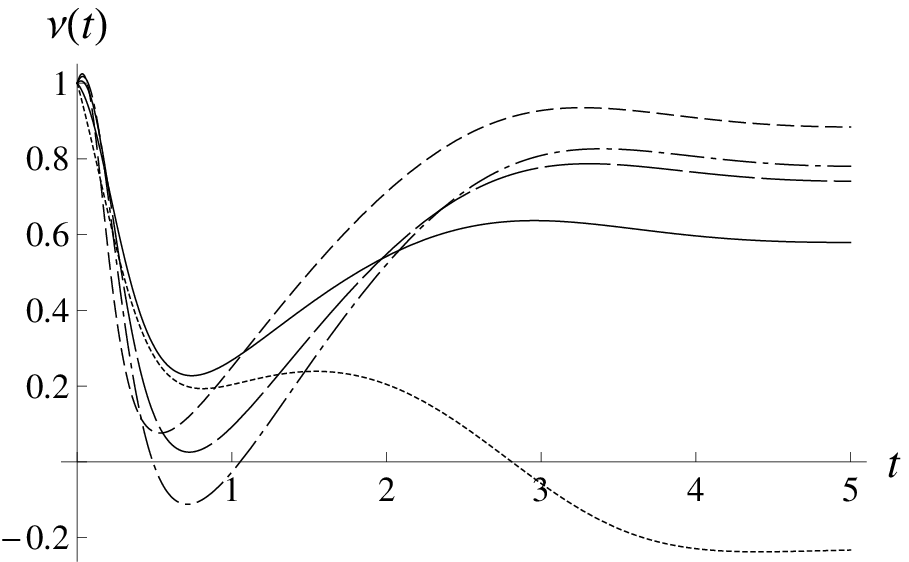}
\includegraphics[width=3.2in]{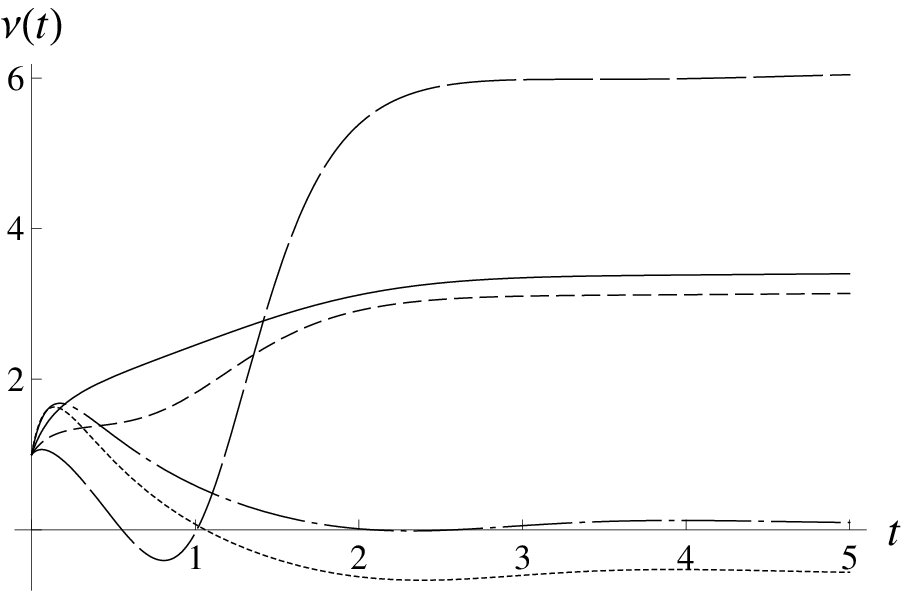}
\caption{(Left) Hu-Paz-Zhang (HPZ) equation. $K_\HPZ$ with initial conditions $\mu_0=\nu_0=1, \kap_0=1$ and coefficients $\th_0=2, \th_2=-\gam, \eta_0=\eta_1=-2\gam b, \gam=1, b=1$. The dotted, solid, long-dashed and dot-dashed curves label $K_\HPZ$  with $\th_1=0.5$, and $\eta_2=1, 0, -1, -1.5$, respectively. The threshold occurs at $\eta_2=0.875$, where $\nu(t)$ approaches 0 in the limit $t\rightarrow \infty$. The short-dashed curve labels $\th_1=-0.5$ and $\eta_2=-1.5$.\newline
(Right) Conjugate of Hu-Paz-Zhang equation. $K_\text{cHPZ}$ with initial conditions $\mu_0=\nu_0=1, \kap_0=1$ and coefficients $\th_0=2, \th_1=0.5, \th_2=\gam, \eta_0=-\eta_1=-2\gam b, \gam=1, b=1$. The dotted, dot-dashed, solid, and long-dashed  curves have $\eta_2=-3, -2, 3, 7$, respectively. The short-dashed curve labels $\th_1=-0.5, \eta_2=7$. The threshold occurs at $\eta_2=-2.125$.}
\label{fig1}
\end{figure}
%%%

Let us define a unitary transformation that maps the coordinates $(Q, P)$ into $(P, -Q)$ (this is a canonical transformation if $Q, P$ are classical phase space coordinates \cite{Goldstein}), together with inversions in the space components $i=1, 2,$ of the vectors $\thv, \ev$. A generator $K(Q, P, \gam, \thv, \ev)$ that undergoes this transformation is denoted by
%%%
\begin{align}   \label{uK}
    K'(Q, P, \gam, \thv, \ev)=K(P, -Q, \gam, \thv', \ev')\,,
\end{align}
%%%
where $\thv'=(\th_0, -\th_1, -\th_2)$, and etc. We find that
%%%
\begin{align}   \label{K'HPZ}
    K'_\HPZ(Q,P,\gam,\thv,\ev)=K_\text{cHPZ}(Q, P, \gam, \thv, \ev)\,.
\end{align}
%%%
Hence, $K_\text{cHPZ}$ is the image of $K_\HPZ$ under this transformation in the Wigner representation. However, they are not physically equivalent because an exchange between $P$ and $Q$ in the Wigner representation will result in an exchange between $Q$ and $r$ in the space coordinates, which exchanges the roles of the probability component and the correlation component of the density matrices.

In \Fig{fig1}, we illustrate the behaviors of the HPZ equation and its conjugate. The left plot of \Fig{fig1} depicts the evolution of $\nu(t)$ for $K_\HPZ$ with the initial conditions $\mu_0=\nu_0=1, \kap_0=1$ and the coefficients $\th_0=2, \th_2=-\gam, \eta_0=\eta_1=-2\gam b, \gam=1, b=1$. The dotted, solid, long-dashed and dot-dashed curves label $K_\HPZ$  with $\th_1=0.5$, and $\eta_2=1, 0, -1, -1.5$, respectively. We find that the positivity of the stationary states improves as $\eta_2$ reduces from 1 across the threshold of 0.875 down to $-1.5$. At the threshold, $\nu(t)$ approaches 0 in the limit $t\rightarrow \infty$, cf.~\Eq{IAG2}. This behavior is consistent with \Eq{IBG2}. Though a large magnitude of $\eta_2$ is favourable for the positivity of the stationary states, however, the positivity in the early period of the evolution deteriorates. This is illustrated by the dot-dashed curve. A large $|\eta_2|$ also tend to violate the uncertainty principle \cite{Dekker84}. On the other hand, the short-dashed curve shows that a negative $\th_1=-0.5$ for $\eta_2=-1.5$ saves the positivity of its evolution. We conclude that large $|\eta_2|$ leads to negative evolution.

In the right plot of \Fig{fig1}, we simulate the evolution of $\nu(t)$ for $K_\text{cHPZ}$ with initial conditions $\mu_0=\nu_0=1, \kap_0=1$ and coefficients $\th_0=2, \th_1=0.5, \th_2=\gam, \eta_0=-\eta_1=-2\gam b, \gam=1, b=1$. The dotted, dot-dashed, solid, and long-dashed curves have $\eta_2=-3, -2, 3,$ and 7, respectively. The short-dashed curve refers to $\th_1=-0.5, \eta_2=7$, which is now positive. The threshold occurs at $\th_2=-2.125$. The improvement of positivity occurs in the opposite direction in the value of $\eta_2$ compared to $K_\HPZ$. Large values of $\eta_2$ again tend to destroy the positivity in the time evolution. Close to the threshold, the stationary state is positive but part of the early period of the evolution could be negative, as illustrated by the dot-dashed curve.

\vspace{10pt}

%%%%%%%%%%%%%%%%%%%%%%%%%%
%       Sectionn         %
%%%%%%%%%%%%%%%%%%%%%%%%%%
%\subsection{Case II: $\eta_2=0$.}
%\label{SecII}
\begin{enumerate}[(I)]
\setcounter{enumi}{1}
\item $\eta_2=0$.

If $\eta_0=0$, then \Eq{diff+decoh} requires $\eta_1=0$. We find that well-behaved probability distribution functions cannot exist since \Eq{hermcond} or \eqref{G0-G1} cannot be fulfilled. Consequently, we must consider $\eta_0\neq 0$.

There are three classes to consider. For $\eta_1\neq0$, we can consider class (IIA) with $\th_2\neq0$ and class (IIB) with $\th_2=0$, whereas for $\eta_1=0$, we consider class (IIC).
\begin{enumerate}[(A)]
\item $\eta_1\neq0$ and $\th_2\neq 0$.

The factorized condition \eqref{G2=0} becomes
%%%
\begin{align}   \label{IIAsep}
-\th_0(\th_2\eta_0+\gam\eta_1)+\th_1(\gam\eta_0+\th_2\eta_1)=0\,.
\end{align}
%%%
Since $\th_0, \th_1$ are independent, we could have either $\th_1\neq0$ or $\th_1= 0$. For $\th_1\neq 0$, the factorized condition \eqref{IIAsep} gives $\eta_1=-\eta_0 \th_2/\gam=-\eta_0\gam/\th_2$, which permits only the solutions $\th_2=-\gam$ or $\gam$, whereas for $\th_1=0$, the factorized condition yields $\eta_1=-\eta_0 \th_2/\gam$. There are altogether three possibilities.

\begin{enumerate}[{(a)}]
\item $\th_1\neq 0, \th_2=-\gam, \eta_1=\eta_0$. CL equation.

The positive semidefinite conditions \eqref{G0-G1} and \eqref{Svsqasymp} yield
%%%
\begin{align}   \label{IIA3G0G1}
    \Gam_0-\Gam_1 > 0: \qquad &\eta_0<0\,,\\
    -\Sv^2\geq 1: \qquad  &\gam \sqrt{\frac{\th_0+\th_1}{\th_0-\th_1}}\leq |\eta_0|\,.
\end{align}
%%%
This case gives $\eta_0+\eta_1=2\eta_0<0$ and $-\eta_0+\eta_1=0$, i.e., the $\d^2/\d Q^2$ diffusion-like term does not appear in the master equation. This gives the CL equation for quantum Brownian motion,
%%%
\begin{align}   \label{CLQP}
    K_\text{CL}(Q,P)=\frac{\th_0}{2}\left(P\frac{\d}{\d Q}-Q\frac{\d}{\d P}\right) -\frac{\th_1}{2}\left(P\frac{\d}{\d Q}+Q\frac{\d}{\d P}\right) -\gam \frac{\d}{\d P}P +\frac{\eta_0}{2}\frac{\d^2}{\d P^2}\,.
%    K_\text{CL}=-\frac{1}{2}(\th_0-\th_1)i\frac{\d^2}{\d Q\d r}+\frac{1}{2}(\th_0+\th_1)iQr +\gam r\frac{\d}{\d r} -\frac{\eta_0}{2}r^2\,.
\end{align}
%%%

\item $\th_1\neq 0, \th_2=\gam, \eta_1=-\eta_0$. Conjugate of CL equation.

The positive semidefinite conditions \eqref{G0-G1} and \eqref{Svsqasymp} require
%%%
\begin{align}   \label{IIA2G0G1}
    \Gam_0-\Gam_1 > 0: \qquad &\eta_0<0\,,\\
    -\Sv^2\geq 1: \qquad &\gam \sqrt{\frac{\th_0-\th_1}{\th_0+\th_1}}\leq |\eta_0|\,.
\end{align}
%%%
This case gives $-\eta_0+\eta_1=-2\eta_0>0$ and $\eta_0+\eta_1=0$, i.e., the decoherence $r^2$ term does not appear in the master equation. In the Wigner representation, we have
%%%
\begin{align}   \label{cCL}
    K_\text{cCL}(Q,P)=\frac{\th_0}{2}\left(P\frac{\d}{\d Q}-Q\frac{\d}{\d P}\right) -\frac{\th_1}{2}\left(P\frac{\d}{\d Q}+Q\frac{\d}{\d P}\right) -\gam\frac{\d}{\d Q}Q +\frac{\eta_0}{2}\frac{\d^2}{\d Q^2}\,.
%    K_\text{dCL}=-\frac{1}{2}(\th_0-\th_1)i\frac{\d^2}{\d Q\d r}+\frac{1}{2}(\th_0+\th_1)iQr -\gam\frac{\d}{\d Q}Q +\frac{\eta_0}{2}\frac{\d^2}{\d Q^2}\,.
\end{align}
%%%
$K_\text{cCL}$ is the image of $K_\CL$ under the unitary transformation defined in \Eq{uK} in the Wigner representation,
%%%
\begin{align}   \label{K'CL}
    K'_\CL(Q,P,\gam,\thv,\ev)=K_\text{cCL}(Q, P, \gam, \thv, \ev)\,.
\end{align}
%%%

%%%
\begin{figure}[t]
  \centering
\includegraphics[width=3.2in]{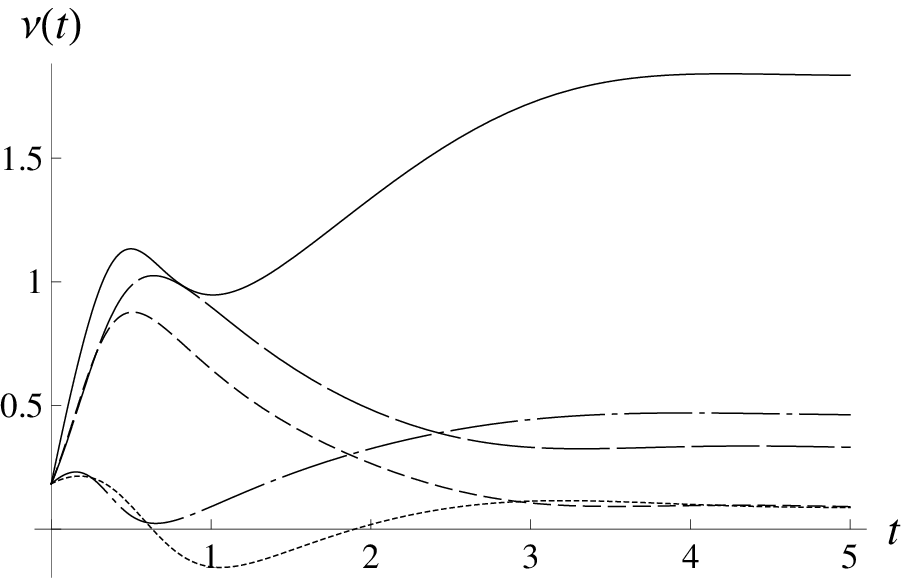}
\includegraphics[width=3.2in]{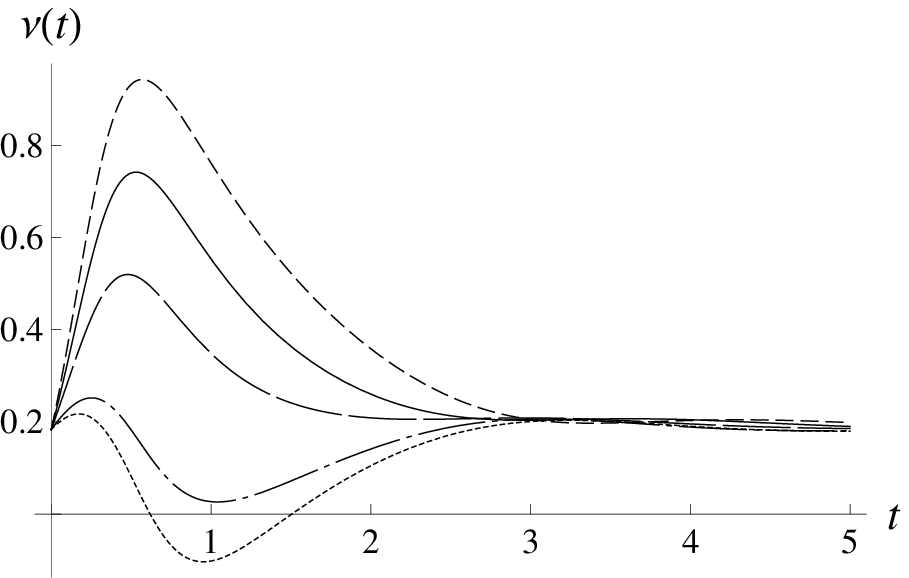}
\caption{(Left) Caldeira-Leggett (CL) equation and its conjugate equivalent. $K_\CL$ and $K_\text{cCL}$ with initial conditions $4\mu_0=1/b_0, \mu_0+\nu_0=b_0=0.6, \kap_0=1$ and coefficients $\th_0=2, \eta_2=0, \gam=1$. Solid, dotted, and dot-dashed curves are $K_\CL$ with $\th_2=-\gam, \eta_0=\eta_1=-2\gam b_\CL$, and $(\th_1, b_\CL)=(0.2, 2), (0.2, 0.6)$, and $(-0.2, 0.6)$, respectively. Long-dashed and short-dashed curves are $K_\text{cCL}$ with $\th_2=\gam, \eta_0=-\eta_1=-2\gam b_\text{CL}$, and $(\th_1, b_\text{CL})=(0.2, 0.6)$ and $(-0.2, 0.6)$, respectively. \newline
(Right) Generalized Caldeira-Leggett equation. $K_\text{gCL}$ with initial conditions $4\mu_0=1/b_0, \mu_0+\nu_0=b_0=0.6, \kap_0=1$ and coefficients $\th_0=2, \th_1=0, \gam=1, b=0.6, \eta_0=-2\gam b, \eta_1=-\eta_0 \th_2/\gam$. The dotted, dot-dashed, long-dashed, solid and short-dashed curves interpolates from $\th_2=-\gam, -0.86 \gam, -0.553 \gam, 0$, to $\gam$, respectively. The generator is completely positive when $\th_2\geq -0.553\gam$.}
\label{fig2}
\end{figure}
%%%

\item $\th_1= 0, \th_2\neq 0, \eta_1=-\eta_0 \th_2/\gam$.
    Generalized CL equation.

Applying the positive semidefinite conditions \eqref{G0-G1} and \eqref{Svsqasymp}, we obtain
%%%
\begin{align}   \label{IIA1G0G1}
    \Gam_0-\Gam_1 > 0: \qquad &\eta_0<0\,,\\
    -\Sv^2\geq 1: \qquad & \gam\leq |\eta_0|\,.
\end{align}
%%%
At this stage, we need to impose $-\eta_0+\eta_1\geq 0$ \eqref{dQ>=0} and $\eta_0+\eta_1\geq 0$ \eqref{r2<=0} to yield
%%%
\begin{align}
    |\th_2|\leq \gam\,.
\end{align}
%%%
As a result, the Wigner representation of its dissipative part interpolates between the two extremes, i.e., the CL equation \eqref{CLQP} with $\th_1=0, \th_2=-\gam, \eta_1=\eta_0$, to its conjugate \eqref{cCL} with $\th_2=\gam, \eta_1=-\eta_0$,
%%%
\begin{align}   \label{genCL}
    K_\text{gCL}(Q,P)=&\frac{\th_0}{2}\left(P\frac{\d}{\d Q}-Q\frac{\d}{\d P}\right) -\frac{1}{2}(\gam+\th_2) \frac{\d}{\d Q}Q -\frac{1}{2}(\gam-\th_2) \frac{\d}{\d P}P \no &+\frac{\eta_0}{4}\left(1+\frac{\th_2}{\gam}\right) \frac{\d^2}{\d Q^2} +\frac{\eta_0}{4}\left(1-\frac{\th_2}{\gam}\right)\frac{\d^2}{\d P^2} \,.
%    K_\text{gCL}=\frac{\th_0}{2}i\left(-\frac{\d^2}{\d Q\d r}+Qr\right) -\frac{1}{2}(\gam+\th_2) \frac{\d}{\d Q}Q +\frac{1}{2}(\gam-\th_2) r\frac{\d}{\d r} +\frac{\eta_0}{4}\left(1+\frac{\th_2}{\gam}\right) \frac{\d^2}{\d Q^2}-\frac{\eta_0}{4}\left(1-\frac{\th_2}{\gam}\right)r^2\,.
\end{align}
%%%
This equation generalizes the CL equation. It is the image of itself under the unitary transformation \eqref{uK} in the Wigner representation,
%%%
\begin{align}   \label{K'gCL}
    K'_\text{gCL}(Q,P,\gam,\thv,\ev)=K_\text{gCL}(Q, P, \gam, \thv, \ev)\,.
\end{align}
%%%
\end{enumerate}
\end{enumerate}
\end{enumerate}

In \Fig{fig2}, we illustrate the behaviors of the three types of CL equations in this class. The left plot depicts $K_\CL$ and $K_\text{cCL}$ with initial conditions $4\mu_0=1/b_0, \mu_0+\nu_0=b_0=0.6, \kap_0=1$ and coefficients $\th_0=2, \eta_2=0, \gam=1$. The solid, dotted, and dot-dashed curves are $K_\CL$ with $\th_2=-\gam, \eta_0=\eta_1=-2\gam b$, and $(\th_1, b)=(0.2, 1), (0.2, 0.6)$, and $(-0.2, 0.6)$, respectively. When this equation is applied to low temperature, it tends to give rise to unphysical behaviour, because the contribution from the vacuum fluctuation, i.e., the 1/2 term on the RHS of \Eq{b} that is neglected in $b_\text{CL}$ \eqref{bCL}, becomes important \cite{Tameshtit12}. This fact is illustrated in the solid and dotted curves when the temperature reduces from $b=1$ to 0.6. The low temperature curve gives negative evolution in the early interval of its evolution, even though its stationary state is positive. This occurs because of an inconsistent application of the CL equation to the low temperature environment. However, a negative $\th_1=-0.2$ saves the positivity of the evolution in the dotted-dashed curve. The long-dashed and short-dashed curves are $K_\text{cCL}$ with $\th_2=\gam, \eta_0=-\eta_1=-2\gam b$, and $(\th_1, b_\text{CL})=(0.2, 0.6)$ and $(-0.2, 0.6)$, respectively. Both curves show positive evolution.

The right plot of \Fig{fig2} refers to $K_\text{gCL}$ with initial conditions $4\mu_0=1/b_0, \mu_0+\nu_0=b_0=0.6, \kap_0=1$ and coefficients $\th_0=2, \th_1=0, \gam=1, b=0.6, \eta_0=-2\gam b, \eta_1=-\eta_0 \th_2/\gam$. The curves interpolates from $\th_2=-\gam, -0.86\gam, -0.553\gam, 0$, to $\gam$, labeled by the dotted, dot-dashed, solid, long-dashed and short-dashed curves, respectively. The dotted and solid curves give the CL and the KL equation, respectively. The plots show that an increase in the value of $\th_2$ improves the positivity of the evolution. In Section \ref{SecCP}, we show that $K_\text{gCL}$ is completely positive when the constraint \eqref{bgCL} is satisfied.

\begin{enumerate}[{(I)}]
\setcounter{enumi}{1}
\item $\eta_2=0$.
\begin{enumerate}[{(A)}]
\setcounter{enumii}{1}
\item $\eta_1\neq 0$ and $\th_2=0$. Generalized KL equation type 1.

The factorized condition \eqref{G2=0} yields $\eta_1=\eta_0\th_1/\th_0$.  The case of $\th_1=0$ overlaps with the situation $\eta_1=0$ considered in class (IIC). The positive semidefinite conditions \eqref{G0-G1} and \eqref{Svsqasymp} require
%%%
\begin{align}   \label{IIAG0G1}
    \Gam_0-\Gam_1 > 0: \qquad &\eta_0<0\,,\\
    -\Sv^2\geq 1: \qquad & \frac{\gam\th_0}{\sqrt{\th_0^2-\th_1^2}}\leq |\eta_0|\,.\label{IIAG2}
\end{align}
%%%
This class generalizes the KL equation,
%%%
\begin{align}   \label{gKL}
        K_\text{gKL1}(Q,P)=&\frac{\th_0}{2}\left(P\frac{\d}{\d Q}-Q\frac{\d}{\d P}\right) -\frac{\th_1}{2}\left(P\frac{\d}{\d Q}+Q\frac{\d}{\d P}\right) -\frac{\gam}{2}\left(\frac{\d}{\d Q}Q +\frac{\d}{\d P}P \right) \no &+\frac{\eta_0}{4}\left(1-\frac{\th_1}{\th_0}\right) \frac{\d^2}{\d Q^2} +\frac{\eta_0}{4}\left(1+\frac{\th_1}{\th_0}\right) \frac{\d^2}{\d P^2} \,.
\end{align}
%%%
$K_\text{gKL1}$ is the image of itself under the unitary transformation \eqref{uK} in the Wigner representation,
%%%
\begin{align}   \label{K'gKL}
    K'_\text{gKL1}(Q,P,\gam,\thv,\ev)=K_\text{gKL1}(Q, P, \gam, \thv, \ev)\,.
\end{align}
%%%
\end{enumerate}
\end{enumerate}

%%%
\begin{figure}[t]
  \centering
\includegraphics[width=3.2in]{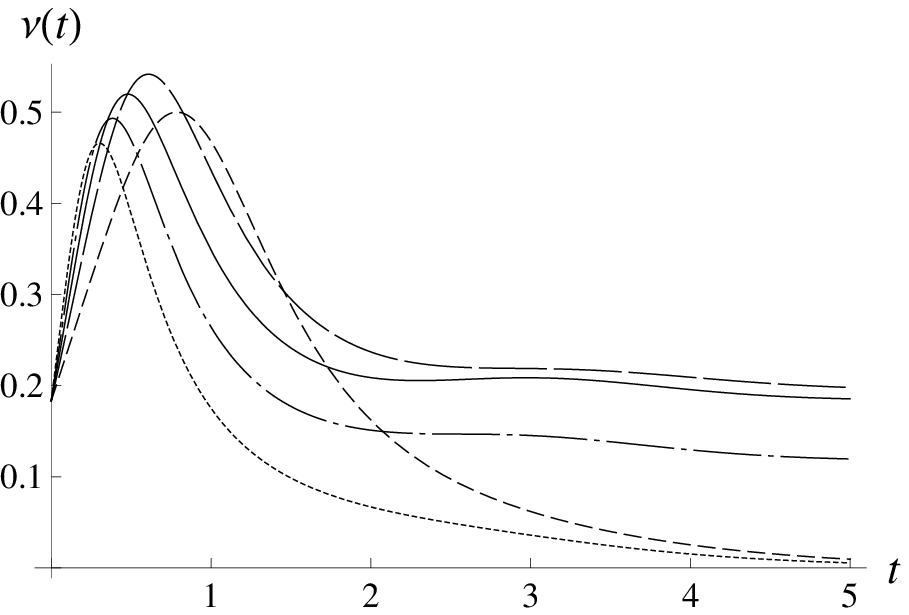}
\includegraphics[width=3.2in]{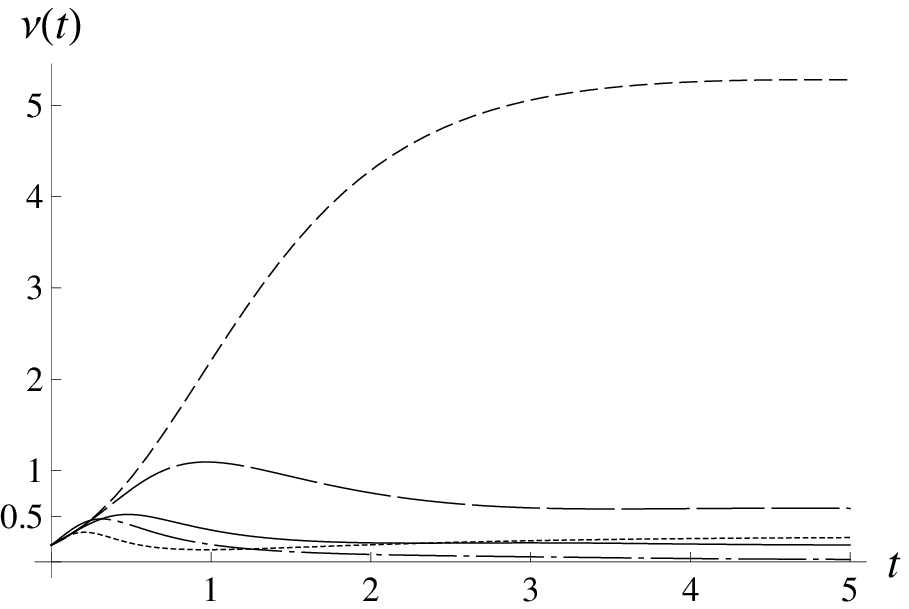}
\caption{(Left) Generalized Kossakowski-Lindblad (KL) equation type 1. $K_\text{gKL1}$ with initial conditions $4\mu_0=1/b_0, \mu_0+\nu_0=b_0=0.6, \kap_0=1$ and coefficients $\th_0=2, \th_2=0, \gam=1, \eta_0=-2\gam b, \eta_1=\eta_0 \th_1/\th_0, \eta_2=0$. The threshold values occur at $\th_1=\pm 1.106$. The dotted, dot-dashed, solid, long-dashed and short-dashed curves label $\th_1=-1.106, -0.5, 0, 0.5$, and 1.106, respectively.\newline
(Right) Generalized Kossakowski-Lindblad equation type 2. $K_\text{gKL2}$ with initial conditions $4\mu_0=1/b_0, \mu_0+\nu_0=b_0=0.6, \kap_0=1$ and coefficients $\th_0=2, \th_2=\gam\th_1/\th_0, \gam=1, \eta_0=-2\gam b, \eta_1=0, \eta_2=0$. The threshold values occur at $\th_1=\pm 1.789$. The dotted, dot-dashed, solid, long-dashed and short-dashed curves label $\th_1=-1.8, -1, 0, 1$, and 1.8, respectively.}
\label{fig3}
\end{figure}
%%%

In the left plot of \Fig{fig3}, we illustrate the behaviors of the generalized KL type 1 class of equations, with initial conditions $4\mu_0=1/b_0, \mu_0+\nu_0=b_0=0.6, \kap_0=1$ and coefficients $\th_0=2, \th_2=0, \eta_0=-2\gam b, \eta_1=\eta_0 \th_1/\th_0, \gam=1, b=0.6$. $K_\text{gKL1}$ interpolates between the threshold values $\th_1=\pm 1.106$, cf.~\eqref{IIAG2}, when $\nu(t)$ approaches 0 in the limit $t\rightarrow \infty$. Outside the threshold values the stationary states are negative. The dotted, dot-dashed, solid, long-dashed and short-dashed curves label $\th_1=-1.106, -0.5, 0, 0.5$, and 1.106, respectively. The solid curve with $\th_1=0$ refers to the KL equation. The plots show that all the evolutions are positive.

\begin{enumerate}[{(I)}]
\setcounter{enumi}{1}
\item $\eta_2=0$.
\begin{enumerate}[{(A)}]
\setcounter{enumii}{2}
\item $\eta_1=0$. Generalized KL equation type 2.

The factorized condition \eqref{SepCond} gives $\th_2=\gam \th_1/\th_0$. When $\th_1=0$, we have $\th_2=0$ to recover the KL equation \eqref{KKL}. The positive semidefinite conditions \eqref{G0-G1} and \eqref{Svsqasymp} yield
%%%
\begin{align}   \label{IICG0G1}
    \Gam_0-\Gam_1 > 0: \qquad &\eta_0<0\,,\\
    -\Sv^2\geq 1: \qquad & \gam\sqrt{1-\frac{\th_1^2}{\th_0^2}}\leq |\eta_0|\,. \label{IICG2}
\end{align}
%%%
If we parameterize $\eta_0$ as $\eta_0=-2\gam b$, then both conditions are always satisfied. In contrast to $K_\text{gKL1}$ in which the diffusion terms are interpolated, now it is the drift terms of $K_\text{gKL2}$ that are interpolated,
%%%
\begin{align}   \label{gKL2}
        K_\text{gKL2}(Q,P)=&\frac{\th_0}{2}\left(P\frac{\d}{\d Q}-Q\frac{\d}{\d P}\right) -\frac{\th_1}{2}\left(P\frac{\d}{\d Q}+Q\frac{\d}{\d P}\right) -\frac{\gam}{2}\left(1+\frac{\th_1}{\th_0}\right)\frac{\d}{\d Q}Q -\frac{\gam}{2}\left(1-\frac{\th_1}{\th_0}\right)\frac{\d}{\d P}P  \no &+\frac{\eta_0}{4} \left(\frac{\d^2}{\d Q^2} +\frac{\d^2}{\d P^2}\right) \,.
\end{align}
%%%
It is the image of itself under the transformation \eqref{uK}
%%%
\begin{align}   \label{K'gKL2}
    K'_\text{gKL2}(Q,P,\gam,\thv,\ev)=K_\text{gKL2}(Q, P, \gam, \thv, \ev)\,.
\end{align}
%%%
\end{enumerate}
\end{enumerate}

The right plot of \Fig{fig3} shows $K_\text{gKL2}$ with the initial conditions $4\mu_0=1/b_0, \mu_0+\nu_0=b_0=0.6, \kap_0=1$ and coefficients $\th_0=2, \th_2=\gam\th_1/\th_0, \gam=1, \eta_0=-2\gam b, \eta_1=0$. When $|\th_1|$ is greater than the threshold value 1.789, the oscillator becomes overdamped. The plots shows the behaviors of $\nu(t)$ when $\th_1$ interpolates from the overdamped region, -1.8 and 1.8 (dotted and short-dashed curves, respectively), into the underdamped region, -1 and 1 (dot-dashed and long-dashed curves, respectively). The solid curve with $\th_1=0$ labels the KL equation. The plots of $\nu(t)$ are all positive as expected for completely positive $K_\text{gKL2}$, a fact which will be shown in Section \ref{SecCP}

We observe that for all the equations from the class (II) and class (IB), $\eta_0<0$ guarantees that $\mu_\text{st}>0$. The only exceptions is the HPZ equation from class (IA), in which $\eta_2$ takes part in ensuring that $\mu_\text{st}>0$. In all the plots, the oscillators are in the underdamped region. For larger damping $\gam$, $\th_2=\pm \gam$ could bring $\w$ into the overdamped region. Due to the way we parameterized $\eta_0=-2\gam b$, the positivity of the evolution is not much affected.

In summary, the HPZ equation and its conjugates, the CL equation, its conjugates and generalizations, and the two types of generalized KL equations (including the KL equation), exhaust the list of equations that are consistent with the reduced dynamics and simultaneously possess stationary states with factorized Wigner functions. Whether the conjugate and generalized equations have any realizations in real system is not known. We summarized the results of this section in Table 1.
%%%
\begin{table}[t]
\center
\begin{tabular}{cccccc}
\label{tab1}
  Class & $\eta_2$ & $\eta_1$ & $\th_2$ & $|\th_1|$ %& $\Gam_0-\Gam_1>0$ & $-\Sv^2\geq 1$
  & Completely positive \\
    \hline\hline
  HPZ (IA) &$\neq0$ &$\eta_0$ &$-\gam$ &$<\th_0$ & no\\
  Conjugate HPZ (IB) &$\neq0$ &$-\eta_0$ &$\gam$ &$<\th_0$ & no\\
  \hline
  CL (IIAa) &0 &$\eta_0$ &$-\gam$ &$<\th_0$ & no\\
  Conjugate CL (IIAb) &0 &$-\eta_0$ &$\gam$ &$<\th_0$ & no\\
  Generalized CL (IIAc) &0 &$\displaystyle -\eta_0\frac{\th_2}{\gam}$ &$|\th_2|\leq \gam$ &0 & $\displaystyle \frac{\gam}{\sqrt{1-\th_2^2/\gam^2}}\leq |\eta_0|$\\
  \hline
  Generalized KL 1 (IIB) &0 &$\displaystyle\eta_0\frac{\th_1}{\th_0}$ &0 &$<\th_0$ & yes\\
  Generalized KL 2 (IIC) &0 &0 &$\displaystyle\gam\frac{\th_1}{\th_0}$ &$<\th_0$ & yes\\
\end{tabular}
\caption{Different classes of master equations with stationary states that satisfy the factorized condition \eqref{G2=0}. In all the classes, $\gam>0$, $\th_0=2\w_0$ and $\eta_0<0$. (IIB) and (IIC) reduce to the KL equation at $\th_1=0$.}
\end{table}
%%%

%%%%%%%%%%%%%%%%%%%%%%%%%%
%       Sectionn         %
%%%%%%%%%%%%%%%%%%%%%%%%%%
\section{Discussions}
\label{SecDisc}

%%%%%%%%%%%%%%%%%%%%%%%%%%
%       Sectionn         %
%%%%%%%%%%%%%%%%%%%%%%%%%%
\subsection{Gibbs states as stationary states}
\label{SecGibbsSt}

After imposing the factorized condition, if we further require $\Gam_1=0$, then we obtain an even smaller subset of master equations whose stationary states are the Gibbs states. All of them will have $\Gam_0=-\eta_0/\gam$, or $\Gam_0=2b$ if we parameterize $\eta_0=-2\gam b$. We list the corresponding master equations below.

Both the HPZ equations (IA) and its conjugate (IB) will be subjected to the additional constraint
%%%
\begin{align}   \label{HPZGibbs}
    \eta_2=\frac{\th_1}{\gam}\eta_0\,.
\end{align}
%%%
If $\th_1=0$, then it reduces to the CL equation \eqref{KCL} with $\th_1=0$.

Both the CL equations (IIAa) and its conjugate (IIAb) should have $\th_1=0$ to ensure that $\Gam_1=0$. In contrast, the generalized CL equations (IIAc) automatically satisfies $\Gam_1=0$. Hence, all the stationary states of the generalized CL equations are the Gibbs states.

The generalized KL equations of type 1 (IIB) and type 2 (IIC) also should have $\th_1=0$. As a result, both of them reduce to the KL equation \eqref{KKL}.

%%%%%%%%%%%%%%%%%%%%%%%%%%
%       Sectionn         %
%%%%%%%%%%%%%%%%%%%%%%%%%%
\subsection{Completely positive generators}
\label{SecCP}

The plots in \Fig{fig3} suggest that $K_\text{gKL1}$ and $K_\text{gKL2}$ are positive operators. We can show that this is indeed true by writing $K_\text{gKL1,2}$ in the form of completely positive operators, which have the generic form
%%%
\begin{subequations}
\begin{align}   \label{CPK}
    K_\text{CP}&=-\sum_i \left(2V^\dg_i\rho V_i-V_i V^\dg_i\rho-\rho V_i V^\dg_i\right)\,, \\
    V_i&=c_i a+d_i \adg\,,
\end{align}
\end{subequations}
%%%
where $a$ and $\adg$ are the annihilation and creation operators of the oscillator, $c_i$ and $d_i$ are complex coefficients. \Eq{CPK} can be written in the form $K_\text{CP}=\gam (O_0-1/2)+\eta_0 O_++\eta_1 L_{1+}+\eta_2 L_{2+}$, provided
%%%
\begin{subequations}
\begin{align}   \label{gamgKL}
    \gam&=-2\sum_i (|c_i|^2-|d_i|^2)\,,\\
    \eta_0&=-2\sum_i (|c_i|^2+|d_i|^2)\,,\\
    \eta_1&=-2\sum_i (c^*_i d_i+c_i d^*_i)\,,\\
    \eta_2&=-2i\sum_i (c^*_i d_i-c_i d^*_i)\,. \label{eta2gKL}
\end{align}
\end{subequations}
%%%
A possible realization of $c_i$ and $d_i$ is
%%%
\begin{align}   \label{cdKL}
    c_i=c\,, \qquad d_i=c s_i\,, \qquad i=1,2\,,
\end{align}
%%%
where $c, s_i$ are real. Solving for $c$ and $s_i$ from \Eqs{gamgKL}-\eqref{eta2gKL}, we find that
%%%
\begin{subequations}
\begin{align}   \label{csKL}
    c&=\frac{1}{8}\sqrt{|\eta_0|-\gam}\,,\\
    s_1&=\frac{1}{|\eta_0|-\gam}\left(-\eta_1+ \sqrt{\eta_0^2-\eta_1^2-\gam^2}\right)\,,\\
    s_2&=\frac{1}{|\eta_0|-\gam}\left(-\eta_1- \sqrt{\eta_0^2-\eta_1^2-\gam^2}\right)\,.
\end{align}
\end{subequations}
%%%
Since $c, s_i$ are real, the expressions under the square roots must be positive semidefinite, which lead to the inequalities,
%%%
\begin{align}   \label{posgKL}
    \gam\leq |\eta_0|\,, \qquad \gam^2 \leq \eta_0^2-\eta_1^2\,.
\end{align}
%%%

For $K_\text{gKL1}$, substituting $\eta_1=\eta_0 \th_1/\th_0$ into the second inequality of \Eq{posgKL} returns the positive condition in \Eq{IIAG2}. Therefore, real solutions always exist and $K_\text{gKL1}$ is completely positive. For $K_\text{gKL2}$ with $\eta_1=0$, it is completely positive provided $\gam\leq |\eta_0|$, which is always true if we parameterize $\eta_0=-2\gam b$.

We can carry out the same consideration on $K_\text{gCL}$. We find that for$\eta_1=-\eta_0 \th_2/\gam$, real solutions exist provided $|\th_2|<\gam$. If we parameterize $\eta_0=-2\gam b$, then there always exist real solutions at high enough temperature, or large $b$, that fulfill the constraint,
%%%
\begin{align}   \label{bgCL}
    \frac{1}{\sqrt{1-\th_2^2/\gam^2}}\leq 2b\,.
\end{align}
%%%
When $\th_2=\pm \gam$, \Eq{bgCL} cannot be fulfilled with finite $b$. Hence, $K_\text{gCL}$ is completely positive within the constraint \eqref{bgCL}. We should emphasize that being a not completely positive generator does not rule out the possibility of the master equation to generate positive evolution for a given initial state provided the coefficients of the master equation are chosen appropriately. This is illustrated in the right plot of \Fig{fig2}. The threshold value of $\th_2$ within which $K_\text{gCL}$ is completely positive is $|\th_2|\leq 0.553\gam$. However, the dot-dashed curve with a value of $\th_2=0.86\gam$ lying outside the threshold still gives positive evolution, although it is not completely positive.

It is impossible for $K_\CL, K_\text{cCL}, K_\HPZ$ and $K_\text{cHPZ}$ to satisfy the second inequality of \Eq{posgKL}. For these generators, $\eta_1=\pm\eta_0$, \Eq{posgKL} could be satisfied if $\gam=0$, in which case we have the trivial case of a free oscillator.

%%%%%%%%%%%%%%%%%%%%%%%%%%
%       Sectionn         %
%%%%%%%%%%%%%%%%%%%%%%%%%%
\subsection{Dissipation caused by unitary components of the generator}
\label{SecU}

We note that even though $iM_2$ belongs to the unitary part of the dynamics, i.e. $\exp(-\th_2 iM_2 t)$ is unitary, but it could affect the positivity of the time evolution through the term $(\thv\cdot\ev)\thv$ and $\thv\wedge\ev$ in $\gv(t)$, cf.~\Eq{gvt}. Mathematically, this is because though the sets of operators $J_0= \{iL_0, iM_1, iM_2\}$ and $J_+=\{O_+, L_{1+}, L_{2+}\}$ are closed separately under the commutator brackets, but $[J_0, J_+]\in J_+$ \cite{Tay16}. Hence, the coefficients of $J_0$ could influence the dissipative part of the dynamics even though they belong to the unitary part.

For a completely different reason, the operator $\exp(v O_0)$ with real $v$ is unitary \cite{Umezawa75,Umezawa75b,Tay07,Tay16} but it causes dissipation because it is not factorizable \cite{Tay11,Tay16}. As a consequence, it maps pure states into mixed states. For example, the oscillator ground state $|0\>\<0|$ is mapped by $\exp(v O_0)$ into the Gibbs state with finite temperature, provided that $\tanh(v/2)=(2b-1)/(2b+1)$ \cite{Tay11}.

%%%%%%%%%%%%%%%%%%%%%%%%%%
%       Sectionn         %
%%%%%%%%%%%%%%%%%%%%%%%%%%
\section{Conclusion}
\label{SecConclusion}

We obtain the generic solution to the most general bilinear master equation with constant coefficients for a quantum oscillator in the form of Gaussian. The properties of the stationary states are determined by the components of a three-dimensional vector in the Minkowski space. We show that a factorized condition on the Wigner function of the stationary states is sufficient to generate a generic class of master equations that includes the well-known ones as special cases. In addition to this, it also generates their conjugates and generalizations. We also show that the generalized KL equations and some of the generalized CL equations are completely positive. For master equations that do not have completely positive generators, although positive semidefinite stationary states are not sufficient to warrant a positive evolution for a given initial state, they serve as references to identify the coefficients of the master equations that are able to generate positive evolution.

\hfill

%%%%%%%%%%%%%%%%%%%%%%%%%%
% Acknowledgmentss %
%%%%%%%%%%%%%%%%%%%%%%%%%%
\noindent \textbf{Acknowledgments}

\hfill

We thank Professor Sujin Suwanna, Dr. Fattah Sakuldee and the Department of Physics, Faculty of Science for hospitality and interesting discussions during our visit at the University of Mahidol, Bangkok, Thailand. This work is supported by the Ministry of Higher Education Malaysia (MOHE) under the Fundamental Research Grant Scheme (FRGS), Project No.~FRGS/2/2014/ST02/UNIM/02/1.

\appendix

%%%%%%%%%%%%%%%%%%%%%%%%%%
%       Sectionn         %
%%%%%%%%%%%%%%%%%%%%%%%%%%
\section{Matrix representation of the time evolution operator}
\label{SecDecomp}

We denote an anti-commutator bracket between $X, Y$ by the notation
%%%
\begin{align}   \label{AXY}
    A_X(Y)\equiv \half\{X,Y\}=\half(XY+YX)\,, \qquad
    A^n_X(Y)=\underbrace{A_X(A_X(\cdots A_X}_\text{$n$-terms}(Y)))\,.
\end{align}
%%%
The identity operator that accompanies $O_0$ gives rise to an overall factor $\exp(\gam t/2)$. For simplicity, we extract it from $K$ and consider $K'=K+\gam/2$ \eqref{Kp}. We can then rewrite the time evolution operator as
%%%
\begin{align}   \label{expL}
    e^{-t\Km' }&= 1 -t\Km'  +\frac{t^2}{2!}\Km'^2+\cdots+\frac{(-t)^n}{n!}\Km'^n+\cdots\\
        &=1-tA_{\Km'}(1)+\frac{t^2}{2!}A_{\Km'}^2(1)+\cdots+\frac{(-t)^n}{n!}A_{\Km'}^n(1)+\cdots\,,
\end{align}
%%%
since
%%%
\begin{align}   \label{AL}
    A_{\Km'}\left({\Km'}^{n-1}\right)&= \half\left\{\Km',{\Km'}^{n-1}\right\}={\Km'}^n \,.
\end{align}
%%%

In terms of the $M_0\equiv L_0$ and $M_\pm\equiv M_1\pm iM_2$ operators, where $\u{M}_\pm^2=\u{0}$, $\Km_0$ then takes the form
%%%
\begin{align}   \label{K0al}
    \Km_0&=i\th_0M_0 +\half (i\th_1+\th_2)M_+ +\half (i\th_1-\th_2)M_-\,.
\end{align}
%%%
Consequently, we deduced that
%%%
\begin{subequations}
\begin{align}   \label{AK0}
    A^0_{\Ku'}(\u{I})&\equiv\u{I}\,,&\qquad
    A^1_{\Ku'}(\u{I})&=A_{\Ku'}(\u{I})\,,\\
    A^2_{\Ku'}(\u{I})&=\al^2\u{I}+\u{H}\,,&\qquad
    A^3_{\Ku'}(\u{I})&=\al^2 A_{\Ku'}(\u{I})+A_{\Ku'}(\u{H})\,,\\
    A^4_{\Ku'}(\u{I})&=(\al^4+\bt^2)\u{I}+2\al^2\u{H}\,,&\qquad
    A^5_{\Ku'}(\u{I})&=(\al^4+\bt^2)A_{\Ku'}(\u{I})+2\al^2A_{\Ku'}(\u{H})\,,\label{AK5}
\end{align}
\end{subequations}
%%%
and so on, where
%%%
\begin{subequations}
\begin{align}   \label{ALL}
%    A_\Km(\u{\Km})&=\u{\Km}_0^2+\u{\Km}_1^2+\big\{\u{\Km}_0,\u{\Km}_1\big\}=\frac{\al^2}{4}+\u{Z}\,,\\
    \al^2&\equiv(\w^2+\gam^2)/4\,,&\qquad
    \bt^2&\equiv \big\{\Ku_0,\Ku'_1\big\}^2/4=\w^2\gam^2/4\,,\\
    \w^2&\equiv-\th_0^2+\th_1^2+\th_2^2\,,\\
    \u{H}&\equiv \big\{\Ku_0,\Ku'_1\big\}=2\gam \Ku_0\u{O}_0-2i(\thv\cdot\ev) \u{M}_0\u{O}_+\,,&\qquad
    \u{M}_0&\u{O}_+=-\u{M}_1\u{L}_{1+}=-\u{M}_2\u{L}_{2+}\,,\label{Hdef}
    %=2(\u{X}+\u{Y})\,,\\
    %\qquad     \u{X}\equiv \gam \Ku_0 \u{O}_0\,,\qquad     \u{Y}\equiv (\alv\cdot\ev) \u{K}_0\u{O}_+ \,,\\
\end{align}
\end{subequations}
%%%
where we have made use of the identities,
%%%
\begin{subequations}
\begin{align}
    4\Ku_0^2&=\w^2\,,\\
    4{\Ku'_1}^2&= \gam^2\,,\\
    A_{\Ku'}(\u{I})&={\Ku'}\,,\label{AIdef}\\
    A_{\Ku'}(\u{H})%&=\u{Q}={\Ku'}(\u{X}+\u{Y})+(\u{X}+\u{Y}){\Ku'}\no
        &=\half\big(\gam^2\Ku_0+\w^2\gam\u{O}_0 +\Theta_0 \u{O}_+ +\Theta_1 \u{L}_{1+} +\Theta_2 \u{L}_{2+}\big)\,,\label{AHdef}\\
    A^2_{\Ku'}(\u{H})&=\bt^2+\al^2\u{H}\,,\\
    \Thv&\equiv (\thv\cdot\ev)\thv+\gam(\thv\wedge\ev)\,.
\end{align}
\end{subequations}
%%%
where $\Thv$ is defined by \Eq{Thv}. Bold face letters denote three-dimensional Minskowski space vectors introduced in Section \ref{Sec4DU}.

\Eqs{AK0}-\eqref{AK5} suggest that $A^n_{\Ku'}(\u{I})$ can be simplified by writing them in terms of $2\times2$ matrices as follows,
%%%
\begin{align}
    A^{2n}_{\Ku'}(\u{I})=\u{A}^n\X\,,\qquad
    A^{2n+1}_{\Ku'}(\u{I})=\u{A}^n\X'\,,\qquad n=0,1,2,\cdots\,,
\end{align}
%%%
where $\u{A}$ is a non-hermitian matrix
%%%
\begin{align}
    \u{A}&\equiv \left(
            \begin{array}{cc} \al^2&\bt^2\\
                                1&\al^2 \end{array} \right)\,,
\end{align}
%%%
and the column matrices denote
%%%
\begin{align}
    \left(\!\!\begin{array}{c} a\\b \end{array} \!\! \right)\equiv a\u{I}+b\u{H}\,,\qquad
    \left(\!\!\begin{array}{c} a\\b \end{array} \!\! \right)'\equiv aA_{\Ku'}(\u{I})+bA_{\Ku'}(\u{H})\,.
\end{align}
%%%
As a result, the time evolution operator can be written as
%%%
\begin{align}   \label{expLW}
    e^{-t\Ku'}&=\sum_{n=0,1,2,\cdots}^\infty
                \left[\frac{(-t)^{2n}}{(2n)!}A_{\Ku'}^{2n}(\u{I})+\frac{(-t)^{2n+1}}{(2n+1)!}A_{\Ku'}^{2n+1}(\u{I})\right]
                =\sum_{n=0,1,2,\cdots}^\infty \left[ \frac{t^{2n}}{(2n)!}\u{A}^n\X-\frac{t^{2n+1}}{(2n+1)!}\u{A}^n\X'\right]\,.
\end{align}
%%%

%%%%%%%%%%%%%%%%%%%%%%%%%%
%       Sectionn         %
%%%%%%%%%%%%%%%%%%%%%%%%%%
%\subsection{Non-hermitian eigenvalue problem}
%\label{SecNonherm}

We can diagonalize this expression by using the biorthogonal basis of $\u{A}$. The right and left eigenvalue problem of the non-hermitian matrix $\u{A}$ are
%%%
\begin{align}   \label{eigenv}
    \u{A}\u{v}_\pm=\lam_\pm\u{v}_\pm\,,\qquad \u{A}^\dg\u{u}_\pm=\lam_\pm\u{u}_\pm\,,
\end{align}
%%%
respectively. The eigenvalues are
%%%
\begin{align}   \label{lambda}
    \lam_\pm&=\al^2\pm \bt=\frac{1}{4}(\gam\pm\w)^2\,,
\end{align}
%%%
and the right and left eigenvectors are
%%%
\begin{align}
    \u{v}_\pm&=\frac{1}{\sqrt{2} \bt} \left(\!\!\begin{array}{c}
                                \bt^2\\
                                \lam_\pm-\al^2 \end{array}\!\! \right)
    =\frac{1}{\sqrt{2}} \left(\!\!\begin{array}{c}
                                \bt\\
                                \pm 1 \end{array}\!\! \right)\,,\qquad
    \u{u}_\pm=\frac{1}{\sqrt{2} \bt} \left(\!\!\begin{array}{c}
                                1\\
                                \lam_\pm-\al^2 \end{array}\!\! \right)
    =\frac{1}{\sqrt{2} } \left(\!\!\begin{array}{c}
                                1/\bt\\
                               \pm 1 \end{array}\!\! \right)\,,
\end{align}
%%%
respectively. The eigenvectors form a pair biorthogonal basis
%%%
\begin{align}   \label{biortho}
    \u{u}_\pm^\dg\cdot\u{v}_\pm=1\,, \qquad  \u{u}_\pm^\dg\cdot\u{v}_\mp=0\,.
\end{align}
%%%
We can form two matrices
%%%
\begin{align}   \label{UV}
    \u{U}%=\frac{1}{\sqrt{2} \bt}
         %   \left(\!\!\begin{array}{cc}
         %               1 & 1 \\
         %               \lam_+-\al^2 & \lam_--\al^2 \end{array}\!\! \right)
        =\frac{1}{\sqrt{2}}
            \left(\!\!\begin{array}{cc}
                        1/\bt & 1/\bt \\
                        1 & -1 \end{array}\!\! \right)\,,\qquad
    \u{V}%=\frac{1}{\sqrt{2} \bt}
         %   \left(\!\!\begin{array}{cc}
         %               \bt^2 & \bt^2 \\
         %               \lam_+-\al^2 & \lam_--\al^2 \end{array}\!\! \right)
    =\frac{1}{\sqrt{2}}
            \left(\!\!\begin{array}{cc}
                        \bt & \bt \\
                         1  & -1 \end{array}\!\! \right),
\end{align}
%%%
which satisfy the relations
%%%
\begin{align}
    \u{U}^\dg\u{V}=\u{V}\u{U}^\dg=\u{I}\,.
\end{align}
%%%
We can now diagonalize $\u{A}$ through
%i%%
\begin{align}   \label{Wdiag}
    \u{U}^\dg\u{A}\u{V}&=\frac{1}{4}\left(\!\!\begin{array}{cc}
                        \lam_+ & 0\\
                        0 & \lam_- \end{array}\!\! \right)
                        =\left(\!\!\begin{array}{cc}
                        (\gam+\w)^2 & 0\\
                        0 & (\gam-\w)^2 \end{array}\!\! \right).
\end{align}
%%%
As a result,
%%%
%\begin{widetext}
\begin{align}   \label{expLdiag}
    e^{-t\Ku'}&=\u{V}\u{U}^\dg e^{-t\Ku'}\no
        &=\u{V}\sum_{n=0,1,2,\cdots}^\infty
            \left(\!\!\begin{array}{cc}
                \frac{\big(\sqrt{\lam_+}t\big)^{2n}}{(2n)!} & 0\\
                0 & \frac{\big(\sqrt{\lam_-}t\big)^{2n}}{(2n)!} \end{array}\!\! \right) \u{U}^\dg\X
            -\u{V}\sum_{n=0,1,2,\cdots}^\infty \left(\!\!\begin{array}{cc}\frac{1}{\sqrt{\lam_+}}\frac{\big(\sqrt{\lam_+}t\big)^{2n+1}}{(2n+1)!}&0\\
            0 & \frac{1}{\sqrt{\lam_-}}\frac{\big(\sqrt{\lam_-}t\big)^{2n+1}}{(2n+1)!} \end{array}\!\! \right) \u{U}^\dg\X' \no
        &=\u{V}\left(\!\!\begin{array}{cc}
                        \cosh\big(\sqrt{\lam_+}t\big) & 0\\
                        0 & \cosh\big(\sqrt{\lam_-}t\big)\end{array}\!\! \right) \u{U}^\dg\X
                    -\u{V}\left(\!\!\begin{array}{cc}\frac{1}{\sqrt{\lam_+}} \sinh\big(\sqrt{\lam_+}t\big) & 0\\
                    0 & \frac{1}{\sqrt{\lam_-}}\sinh\big(\sqrt{\lam_-}t\big) \end{array}\!\! \right) \u{U}^\dg\X'\no
        &=\half\bigg(\cosh\big[(\gam+\w)t/2\big]+\cosh\big[(\gam-\w)t/2\big]\bigg)\u{I}
                +\frac{1}{\bt}\bigg(\cosh\big[(\gam+\w)t/2\big]-\cosh\big[(\gam-\w)t/2\big]\bigg)\u{H} \no
        &\quad-\half\bigg(\frac{\sinh\big[(\gam+\w )t/2\big]}{\gam+\w}
                +\frac{\sinh\big[(\gam-\w)t/2\big]}{\gam-\w}\bigg)A_{\Ku'}(\u{I})
                -\frac{2}{\bt}\bigg(\frac{\sinh\big[(\gam+\w)t/2\big]}{\gam+\w} -\frac{\sinh\big[(\gam-\w)t/2\big]}{\gam-\w}\bigg)A_{\Ku'}(\u{H})\,.
\end{align}
%\end{widetext}
%%%
After simplifying this expression and making use of \Eqs{Hdef}, \eqref{AIdef} and \eqref{AHdef}, we finally obtain the matrix representation of the time evolution operator through
%%%
\begin{align}
    e^{-t\Ku}&=e^{\gam t/2}e^{-t\Ku'}\,,
\end{align}
%%%
which is given in \Eqs{expL2D}-\eqref{Bv}. Here we obtain the BCH formula for real $\w$ for the overdamped oscillator. After we obtain the solutions to the master equation, we analytically continue the solutions to imaginary $\w$ for the underdamped oscillator.

%%%%%%%%%%%%%%%%%%%%%%%%%%
%       Sectionn         %
%%%%%%%%%%%%%%%%%%%%%%%%%%
\section{Coefficients of the BCH formula}
\label{AppMrepresBCH}

Multiplying the matrix on the RHS of \Eq{Kdecomp4D}, we can simplify the BCH formula to yield
%%%
\begin{align} \label{Kdecomp4Dapp}
    e^{- \Ku t}&= e^{g_2\u{L}_{2+}}e^{g_1\u{L}_{1+}}e^{g_0\u{O}_+}e^{h(\u{O}_0-1/2)}e^{m_+\u{M}_+}e^{\ln m_0\u{M}_0}e^{m_-\u{M}_-}\no
    &=e^{-h/2}\left\{\cosh(h/2)\left(\cosh\big(\ln\!\!\sqrt{m_0}\big)-\frac{m_+m_-}{2\sqrt{m_0}}\right)\u{I}
        +2\sinh(h/2)\left(\cosh\big(\ln\!\!\sqrt{m_0}\big)-\frac{m_+m_-}{2\sqrt{m_0}}\right)\u{O}_0 \right.\no
    &\qquad\qquad
        +\cosh(h/2)\left(2\sinh\big(\ln\!\!\sqrt{m_0}\big)-\frac{m_+m_-}{\sqrt{m_0}}\right)\u{M}_0 +\cosh(h/2)\frac{m_+}{\sqrt{m_0}}\u{M}_+ +\cosh(h/2)\frac{m_-}{\sqrt{m_0}}\u{M}_- \no
    &\qquad\qquad +\sinh(h/2) \left[
        \left(2\sinh\big(\ln\!\!\sqrt{m_0}\big)-\frac{m_+m_-}{\sqrt{m_0}}\right)\u{M}_0\u{O}_0 +\frac{m_+}{\sqrt{m_0}}\u{M}_+\u{O}_0 +\frac{m_-}{\sqrt{m_0}}\u{M}_-\u{O}_0\right]\no
    &\qquad\qquad
        +\left[g_0\left(2\sinh\big(\ln\!\!\sqrt{m_0}\big)-\frac{m_+m_-}{\sqrt{m_0}}\right) -g_1\left(\frac{m_++m_-}{\sqrt{m_0}}\right)-ig_2\left(\frac{m_+-m_-}{\sqrt{m_0}}\right)\right]\u{M}_0\u{O}_+\no
    &\qquad\qquad
        +\left[g_0\left(\cosh\big(\ln\!\!\sqrt{m_0}\big)-\frac{m_+m_-}{2\sqrt{m_0}}\right) +g_1\left(\frac{m_+-m_-}{2\sqrt{m_0}}\right)+ig_2\left(\frac{m_++m_-}{2\sqrt{m_0}}\right)\right]\u{O}_+
        \no
    &\qquad\qquad
        +\left[g_1\left(\cosh\big(\ln\!\!\sqrt{m_0}\big)-\frac{m_+m_-}{2\sqrt{m_0}}\right)  +g_0\left(\frac{m_+-m_-}{2\sqrt{m_0}}\right)+ig_2\left(\sinh\big(\ln\!\!\sqrt{m_0}\big)-\frac{m_+m_-}{2\sqrt{m_0}}\right)\right]\u{L}_{1+}
%        \end{align}
%\begin{align}
    \no
   &\qquad\qquad
        +\left.\left[g_2\left(\cosh\big(\ln\!\!\sqrt{m_0}\big)-\frac{m_+m_-}{2\sqrt{m_0}}\right) +ig_0\left(\frac{m_++m_-}{2\sqrt{m_0}}\right)-ig_1\left(\sinh\big(\ln\!\!\sqrt{m_0}\big)-\frac{m_+m_-}{2\sqrt{m_0}}\right)\right]\u{L}_{2+}\right\}\,,
\end{align}
%\end{widetext}
%%%
where we have omitted the time-dependence of the coefficients to simplify the expression.

Comparing the coefficient of $\u{I}$ in \Eq{Kdecomp4Dapp} and \Eq{expL2D}, we obtain \Eq{ht}. Furthermore, a comparison of the coefficients of $\u{O}_0$, $\u{M}_0$ and $\u{M}_\pm$ gives
%%%
\begin{subequations}
\begin{align}   \label{coeffO0}
    \cosh\big(\ln\!\!\sqrt{m_0}\big)-\frac{m_+m_-}{2\sqrt{m_0}}=\cosh(\w t/2)\,,\\
    \sinh\big(\ln\!\!\sqrt{m_0}\big) -\frac{m_+m_-}{2\sqrt{m_0}}=-i\thh_0\sinh(\w t/2)\,,\label{coeffK0}
\end{align}
\end{subequations}
%%%
and \Eq{coeffKpm}, respectively, where the unit vector $\thvh$ is defined in \Eq{thh}. \Eqs{coeffK0} and \eqref{coeffKpm} are consistent with the coefficients obtained by comparing the coefficients of $\u{M}_0\u{O}_0, \u{M}_+\u{O}_0$ and $\u{M}_-\u{O}_0$ in \Eq{Kdecomp4Dapp} and \Eq{expL2D}. \Eqs{coeffO0}-\eqref{coeffK0} can be solved for $m_0$ to give \Eq{A0}. We note that the solutions to $m_0, m_\pm$ are the same as those obtained in Ref.~\cite{Ban93}.

Substituting \Eqs{coeffK0} and \eqref{coeffKpm} into the coefficients of $\u{O}_+, \u{L}_{1+}$ and $\u{L}_{2+}$ in \Eq{Kdecomp4Dapp}, and equating them to the coefficients of \Eq{expL2D}, we obtain
%%%
\begin{align}   \label{coeffOL}
    e^{\gam t/2}\big[ \cosh(\w t/2)\gv+2\sinh(\w t/2)(\thvh\wedge\gv)\big]=-\Sigv\,,
\end{align}
%%%
where $\Sigv$ is defined in \Eq{Bv}. \Eq{coeffOL} can be inverted for $\g_0, \g_1$ and $\g_2$ to yield \Eq{gvt}.

Comparing the coefficient of $\u{M}_0\u{O}_+$ in \Eq{Kdecomp4Dapp} and \Eq{expL2D}, we obtain
%%%
\begin{align}   \label{coeffO+K0}
    \thv\cdot\gv e^{\gam t/2}=-\thv\cdot\ev\frac{\sinh(\gam t/2)}{\gam/2}\,.
\end{align}
%%%
If we take the dot product of \Eqs{coeffOL} and \eqref{Bv} with $\thv$ and $-\thv$, respectively, and set them equal, we obtain \Eq{coeffO+K0}. Hence, \Eq{coeffO+K0} does not provide new information.

%%%%%%%%%%%%%%%%%%%%%%%%%%
%       Sectionn         %
%%%%%%%%%%%%%%%%%%%%%%%%%%
\section{4D matrix representation of the generators}
\label{AppMrepres}

For completeness, we list three other generators that together  with the set of generators in \Eqs{iL0}-\eqref{O+} form a complete basis, cf.~Ref.~\cite{Tay06},
%%%
\begin{align}   \label{O-}
    O_- = -\frac{\d^2}{\d r^2}+Q^2\,, \qquad
    L_{1-} = \frac{\d^2}{\d r^2}+Q^2\,, \qquad
    L_{2-} = 2iQ\frac{\d}{\d r}\,.
\end{align}
%%%
The generators in \Eq{O-} do not conserve probability of the reduced dynamics. Their 4D matrix representations are
%%%
\begin{align}   \label{O-4d}
   \u{O}_- =2 \left(\begin{array}{cc}
            0&\u{\sig}_u\\
            \u{\sig}_d&0
          \end{array}\right), \quad
    \u{L}_{1-} = 2 \left(\begin{array}{cc}
            0&\u{\sig}_u\\
            -\u{\sig}_d&0
          \end{array}\right), \quad
     \u{L}_{2-} = 2i\left(\begin{array}{cc}
            \u{\sig}_+&0\\
            0&-\u{\sig}_-
          \end{array}\right).
\end{align}

The matrix representation of the set of generators \eqref{genQr} can be extracted using \Eq{SJ} directly or by solving \Eqs{L04d}-\eqref{O+d} and \eqref{O-4d} for them. The results are
%%%
\begin{subequations}
\begin{align}   \label{JQr}
    \u{J}\left(\half Q^2\right) &= \left(\begin{array}{cc}
                0&\u{\sig}_u\\
                0&0
          \end{array}\right),
        \quad
            \u{J}(iQr) = i\left(\begin{array}{cc}
            0&\u{\sig}_1\\
            0&0
          \end{array}\right),
            \quad
        \u{J}\left(\half r^2\right) = \left(\begin{array}{cc}
            0&\u{\sig}_d\\
            0&0
          \end{array}\right),
\end{align}
%%%
%%%
\begin{align}
    \u{J}\left(\half \frac{\d^2}{\d Q^2}\right) = \left(\begin{array}{cc}
            0&0\\
            -\u{\sig}_u&0
          \end{array}\right),   \quad
    \u{J}\left(i\frac{\d^2}{\d Q \d r}\right) = i\left(\begin{array}{cc}
            0&0\\
            -\u{\sig}_1&0
          \end{array}\right),   \quad
    \u{J}\left(\half \frac{\d^2}{\d r^2}\right) = \left(\begin{array}{cc}
            0&0\\
            -\u{\sig}_d&0
          \end{array}\right),
\end{align}
%%%
%%%
\begin{align}
    \u{J}\left(iQ\frac{\d}{\d r}\right) = i\left(\begin{array}{cc}
            \u{\sig}_+&0\\
            0&-\u{\sig}_-
          \end{array}\right),\quad
    \u{J}\left(ir\frac{\d}{\d Q}\right) = i\left(\begin{array}{cc}
            \u{\sig}_-&0\\
            0&-\u{\sig}_+
          \end{array}\right),
\end{align}
%%%
%%%
\begin{align}
    \u{J}\left(Q\frac{\d}{\d Q} +\half\right) &= \left(\begin{array}{cc}
                \u{\sig}_u&0\\
                0&-\u{\sig}_u
          \end{array}\right),\quad
    \u{J}\left(r\frac{\d}{\d r} +\half\right) = \left(\begin{array}{cc}
            \u{\sig}_d&0\\
            0&-\u{\sig}_d
          \end{array}\right).
\end{align}
\end{subequations}
%%%

%%%%%%%%%%%%%%%%%%%%%%%%%%
%       Sectionn         %
%%%%%%%%%%%%%%%%%%%%%%%%%%
\section{Proof of \Eqs{evomu}-\eqref{C} as solutions to equations of motion}
\label{AppEqMotProof}

Starting from the expression of $D(t)$ in \Eq{Dt}, we take the time derivative to obtain
%%%
\begin{align}   \label{Ddot}
    \dt{D}&=2\mu_0 \gam e^{\gam t}(\g_0-\g_1) +2\mu_0 e^{\gam t}\left(\dt{\g_0}-\dt{\g_1}\right) +\frac{\w}{2}e^{\w t}(1+C'-B')-\frac{\w}{2}e^{-\w t}(1+C'+B')\no
    &=-2\mu_0e^{\gam t}\big[\eta_0-\eta_1+(\thv\wedge\gv)_0-(\thv\wedge\gv)_1\big] +\frac{\w}{2}e^{\w t}(1+C'-B')-\frac{\w}{2}e^{-\w t}(1+C'+B')\,,
\end{align}
%%%
where we substitute the time derivative of $g_i$ using \Eq{dgdt}. Then, we obtain
%%%
\begin{align}   \label{mudotLHS}
    \dt{\mu}&=\frac{1}{D}\dt{\mu'}-\frac{\mu}{D}\dt{D}=\gam \mu-\frac{\mu}{D}\dt{D}\,.
\end{align}
%%%
Comparing \Eq{mudotLHS} with \Eq{ratemu}, we find that
%%%
\begin{align}   \label{mudotRHS}
    \dt{D}=-(\th_0-\th_1)\kap' -\th_2 D -2\mu_0e^{\gam t}(\eta_0-\eta_1)\,.
\end{align}
%%%
The proof is therefore reduced to showing that the coefficients of the various exponents, i.e., $1, \exp(\gam t), \exp(\pm\w t)$, on the RHS of \Eqs{Ddot} and \eqref{mudotRHS} are equal. In the proof, we convert the terms $g_0+g_1, \thh_0+\thh_1$ into $g_0-g_1, \th_0-\th_1, \thh_2 g_2, (\thvh\wedge\gv)_2, \thvh\cdot\gv$ by using the following identities,
%%%
\begin{align}   \label{th+th}
    (\thh_0+\thh_1)(\thh_0-\thh_1)&=\thh_0^2-\thh_1^2=-1+\thh_2^2\,,\\
    (\thh_0+\thh_1)(g_0-g_1)&=-\thvh\cdot\gv+\thh_2 g_2+(\thvh\wedge\gv)_2\,.
\end{align}
%%%

The equation of motion of $\kap$ gives
%%%
\begin{align}
    \dt{\kap}&=\frac{1}{D}\dt{\kap'}-\frac{\kap}{D}\dt{D}\no
            &=\th_2 \kap+2\mu\eta_2 +2(\eta_0-\eta_1)\mu\kap +2(\thv\wedge\gv)_2 \mu +(\th_0-\th_1)\kap^2 +\frac{\w}{2} \frac{e^{\w t}}{D} (1+C-B) -\frac{\w}{2} \frac{e^{-\w t}}{D}(1+C+B)\,,\label{kdot}
\end{align}
%%%
where we have used \Eq{mudotRHS}. Comparing \Eq{kdot} with \Eq{ratekap}, we need to prove that
%%%
\begin{align}   \label{kdotRHSLHS}
    -\half(\th_0-\th_1)(\del'^2+\kap'^2)+\half(\th_0+\th_1)D^2
    =2\mu_0(\thv\wedge\gv)_2 e^{\gam t}D
            +\frac{\w}{2} e^{\w t} (1+C-B) D
            -\frac{\w}{2} e^{-\w t}(1+C+B)D\,,
\end{align}
%%%
by showing that the coefficients of the various exponents on both sides of \Eq{kdotRHSLHS} are equal. The following identities of the projectors $\Pv_\pm(\ev)$ could be helpful in the proof,
%%%
\begin{align}
    \thvh\wedge\Pv_\pm(\ev)=\mp\Pv_\pm(\ev)\,,\qquad
    \Pv_\pm(\Pv_\mp(\ev))=0\,,\qquad
    \Pv_+(-\ev)=-\Pv_-(\ev)\,.
\end{align}
%%%

Lastly, the equation of motion of $\mu+\nu$ gives
%%%
\begin{align}
    \dt{\mu}+\dt{\nu}&=-\frac{1}{D}\left(\dt{\mu'}+\dt{\nu'}\right)-(\mu+\nu)\frac{1}{D}\dt{D}\no
            &=-\gam(\mu+\nu)+\th_2(\mu+\nu)+(\th_0-\th_1)(\mu+\nu)\kap +2(\eta_0-\eta_1)\mu(\mu+\nu)  \no
            &\quad +2 (\gv\cdot\ev) \mu - \frac{\thvh\cdot\ev}{2D} \Phi(\thvh) -\frac{e^{\w t}}{2D} \Phi\big[\Pv_-(\ev)\big]
                -\frac{e^{-\w t}}{2D} \Phi\big[\Pv_+(\ev)\big]\,. \label{munudot}
\end{align}
%%%
Comparing \Eq{munudot} with \Eq{ratemunu}, we need to prove that
%%%
\begin{align}   \label{munudotLHSRHS}
    &-\half(\eta_0-\eta_1)\kap'^2-\half(\eta_0+\eta_1)D^2-\eta_2\kap' D \no
            &\quad= 2\mu_0(\eta_0-\eta_1)e^{\gam t}(\mu'+\nu') +2\mu_0 (\gv\cdot\ev)e^{\gam t} D -\frac{D}{2}(\thvh\cdot\ev)\Phi(\thvh)
            -\frac{e^{\w t}}{2} \Phi\big[\Pv_-(\ev)\big]D
            -\frac{e^{-\w t}}{2}  \Phi\big[\Pv_+(\ev)\big]D\,.
\end{align}
%%%

%\bibliography{poscohMME}{}

\begin{thebibliography}{10}
\expandafter\ifx\csname url\endcsname\relax
  \def\url#1{\texttt{#1}}\fi
\expandafter\ifx\csname urlprefix\endcsname\relax\def\urlprefix{URL }\fi
\expandafter\ifx\csname href\endcsname\relax
  \def\href#1#2{#2} \def\path#1{#1}\fi

\bibitem{Breuer}
H.-P. Breuer, F.~Petruccione, The Theory of Open Quantum Systems, Oxford, New
  York, 2002.

\bibitem{Gardiner}
C.~W. Gardiner, P.~Zoller, Quantum Noise, 3rd Edition, Springer-Verlag, Berlin,
  2004.

\bibitem{Haake85}
F.~Haake, R.~Reibold, Strong damping and low-temperature anomalies for the
  harmonic oscillator, Phys. Rev. A 32 (1985) 2462--2475.
\newblock \href {http://dx.doi.org/10.1103/PhysRevA.32.2462}
  {\path{doi:10.1103/PhysRevA.32.2462}}.

\bibitem{Diosi93a}
L.~Di\'osi, \href{http://stacks.iop.org/0295-5075/22/i=1/a=001}{On
  high-temperature Markovian equation for quantum Brownian motion}, Euro. Phys.
  Lett. 22 (1993) 1.
\newline\urlprefix\url{http://stacks.iop.org/0295-5075/22/i=1/a=001}

\bibitem{Tameshtit12}
A.~Tameshtit, Zero-point energies, the uncertainty principle, and positivity of
  the quantum Brownian density operator, Phys. Rev. E 85 (2012) 042103.
\newblock \href {http://dx.doi.org/10.1103/PhysRevE.85.042103}
  {\path{doi:10.1103/PhysRevE.85.042103}}.

\bibitem{Suarez92}
A.~Su\'{a}rez, R.~Silbey, I.~Oppenheim, Memory effects in the relaxation of
  quantum open systems, J. Chem. Phys. 97 (1992) 5101--5107.
\newblock \href {http://dx.doi.org/http://dx.doi.org/10.1063/1.463831}
  {\path{doi:http://dx.doi.org/10.1063/1.463831}}.

\bibitem{Munro96}
W.~J. Munro, C.~W. Gardiner, Non-rotating-wave master equation, Phys. Rev. A 53
  (1996) 2633--2640.
\newblock \href {http://dx.doi.org/10.1103/PhysRevA.53.2633}
  {\path{doi:10.1103/PhysRevA.53.2633}}.

\bibitem{Whitney08}
R.~S. Whitney, \href{http://stacks.iop.org/1751-8121/41/i=17/a=175304}{Staying
  positive: going beyond Lindblad with perturbative master equations}, J. Phys.
  A: Math. Theor. 41 (2008) 175304.
\newline\urlprefix\url{http://stacks.iop.org/1751-8121/41/i=17/a=175304}

\bibitem{Fleming10}
C.~Fleming, N.~I. Cummings, C.~Anastopoulos, B.~L. Hu,
  \href{http://stacks.iop.org/1751-8121/43/i=40/a=405304}{The rotating-wave
  approximation: consistency and applicability from an open quantum system
  analysis}, J. Phys. A: Math. Theor. 43 (2010) 405304.
\newline\urlprefix\url{http://stacks.iop.org/1751-8121/43/i=40/a=405304}

\bibitem{Pechukas94}
P.~Pechukas, Reduced dynamics need not be completely positive, Phys. Rev. Lett.
  73 (1994) 1060--1062.
\newblock \href {http://dx.doi.org/10.1103/PhysRevLett.73.1060}
  {\path{doi:10.1103/PhysRevLett.73.1060}}.

\bibitem{Shaji04}
T.~F. Jordan, A.~Shaji, E.~C.~G. Sudarshan, Dynamics of initially entangled
  open quantum systems, Phys. Rev. A 70 (2004) 052110.
\newblock \href {http://dx.doi.org/10.1103/PhysRevA.70.052110}
  {\path{doi:10.1103/PhysRevA.70.052110}}.

\bibitem{Kossa76}
V.~Gorini, A.~Kossakowski, E.~C.~G. Sudarshan, Completely positive dynamical
  semigroups of N-level systems, J. Math. Phys. 17 (1976) 821--825.
\newblock \href {http://dx.doi.org/http://dx.doi.org/10.1063/1.522979}
  {\path{doi:http://dx.doi.org/10.1063/1.522979}}.

\bibitem{Lindblad76}
G.~Lindblad, On the generators of quantum dynamical semigroups, Commun. Math.
  Phys. 48 (1976) 119--130.
\newblock \href {http://dx.doi.org/10.1007/BF01608499}
  {\path{doi:10.1007/BF01608499}}.

\bibitem{Shaji05}
A.~Shaji, E.~C.~G. Sudarshan, Who's afraid of not completely positive maps?,
  Phys. Lett. A 341 (2005) 48 -- 54.
\newblock \href {http://dx.doi.org/10.1016/j.physleta.2005.04.029}
  {\path{doi:10.1016/j.physleta.2005.04.029}}.

\bibitem{Sudarshan61}
E.~C.~G. Sudarshan, P.~M. Mathews, J.~Rau, Stochastic dynamics of
  quantum-mechanical systems, Phys. Rev. 121 (1961) 920--924.
\newblock \href {http://dx.doi.org/10.1103/PhysRev.121.920}
  {\path{doi:10.1103/PhysRev.121.920}}.

\bibitem{Talkner81}
P.~Talkner, Gauss Markov process of a quantum oscillator, Z. Phys. B: Cond.
  Matt. 41 (1981) 365--374.
\newblock \href {http://dx.doi.org/10.1007/BF01307328}
  {\path{doi:10.1007/BF01307328}}.

\bibitem{Tay16}
B.~A. Tay, Symmetry of bilinear master equations for a quantum oscillator,
  Physica A 468 (2017) 578 -- 589.
\newblock \href
  {http://dx.doi.org/http://dx.doi.org/10.1016/j.physa.2016.10.067}
  {\path{doi:http://dx.doi.org/10.1016/j.physa.2016.10.067}}.

\bibitem{Vacchini02}
B.~Vacchini, Quantum optical versus quantum Brownian motion master equation in
  terms of covariance and equilibrium properties, J. Math. Phys. 43 (2002)
  5446--5458.
\newblock \href {http://dx.doi.org/http://dx.doi.org/10.1063/1.1505126}
  {\path{doi:http://dx.doi.org/10.1063/1.1505126}}.

\bibitem{Agarwal71}
G.~S. Agarwal, Brownian motion of a quantum oscillator, Phys. Rev. A 4 (1971)
  739--747.
\newblock \href {http://dx.doi.org/10.1103/PhysRevA.4.739}
  {\path{doi:10.1103/PhysRevA.4.739}}.

\bibitem{Walls85}
D.~F. Walls, G.~J. Milburn, Effect of dissipation on quantum coherence, Phys.
  Rev. A 31 (1985) 2403--2408.
\newblock \href {http://dx.doi.org/10.1103/PhysRevA.31.2403}
  {\path{doi:10.1103/PhysRevA.31.2403}}.

\bibitem{Walls08}
D.~F. Walls, G.~J. Milburn, Quantum Optics, 2nd Edition, Springer-Verlag,
  Berlin, 2008.

\bibitem{Hu11}
C.~Fleming, A.~Roura, B.~Hu, Exact analytical solutions to the master equation
  of quantum Brownian motion for a general environment, Ann. Phys. 326 (2011)
  1207 -- 1258.
\newblock \href {http://dx.doi.org/http://dx.doi.org/10.1016/j.aop.2010.12.003}
  {\path{doi:http://dx.doi.org/10.1016/j.aop.2010.12.003}}.

\bibitem{Tameshtit13}
A.~Tameshtit, On the standard quantum Brownian equation and an associated class
  of non-autonomous master equations, Physica A 392 (2013) 427 -- 443.
\newblock \href
  {http://dx.doi.org/http://dx.doi.org/10.1016/j.physa.2012.09.006}
  {\path{doi:http://dx.doi.org/10.1016/j.physa.2012.09.006}}.

\bibitem{Isar99}
A.~Isar, Uncertainty, entropy and decoherence of the damped harmonic oscillator
  in the Lindblad theory of open quantum systems, Forts. Phys. 47 (1999)
  855--879.
\newblock \href
  {http://dx.doi.org/10.1002/(SICI)1521-3978(199909)47:7/8<855::AID-PROP855>3.0.CO;2-Z}
  {\path{doi:10.1002/(SICI)1521-3978(199909)47:7/8<855::AID-PROP855>3.0.CO;2-Z}}.

\bibitem{Dekker84}
H.~Dekker, M.~Valsakumar, A fundamental constraint on quantum mechanical
  diffusion coefficients, Phys. Lett. A 104 (1984) 67 -- 71.
\newblock \href
  {http://dx.doi.org/http://dx.doi.org/10.1016/0375-9601(84)90964-2}
  {\path{doi:http://dx.doi.org/10.1016/0375-9601(84)90964-2}}.

\bibitem{Giulini}
D.~Giulini, E.~Joos, C.~Kiefer, J.~Kupsch, I.~O. Stamatescu, H.~D. Zeh,
  Decoherence and the Appearance of a Classical World in Quantum Theory,
  Springer, Berlin, 1996.

\bibitem{Zurek91}
W.~H. Zurek, Decoherence and the transition from quantum to classical, Phys.
  Today 44 (1991) 36--44.
\newblock \href {http://dx.doi.org/dx.doi.org/10.1063/1.881293}
  {\path{doi:dx.doi.org/10.1063/1.881293}}.

\bibitem{Klauder68}
J.~R. Klauder, E.~C.~G. Sudarshan, Fundamentals of Quantum Optics, Dover, New
  York, 2006.

\bibitem{Simon87}
R.~Simon, E.~C.~G. Sudarshan, N.~Mukunda, Gaussian-Wigner distributions in
  quantum mechanics and optics, Phys. Rev. A 36 (1987) 3868--3880.
\newblock \href {http://dx.doi.org/10.1103/PhysRevA.36.3868}
  {\path{doi:10.1103/PhysRevA.36.3868}}.

\bibitem{Gilmore}
R.~Gilmore, Lie Groups, Lie Algebras, and Some of Their Applications, John
  Wiley \& Sons, New York, 1974.

\bibitem{Simon89}
R.~Simon, N.~Mukunda, E.~C.~G. Sudarshan, The theory of screws: A new geometric
  representation for the group su(1,1), J. Math. Phys. 30 (1989) 1000--1006.
\newblock \href {http://dx.doi.org/http://dx.doi.org/10.1063/1.528365}
  {\path{doi:http://dx.doi.org/10.1063/1.528365}}.

\bibitem{Wei63}
J.~Wei, E.~Norman, Lie algebraic solution of linear differential equations, J.
  Math. Phys. 4 (1963) 575--581.
\newblock \href {http://dx.doi.org/http://dx.doi.org/10.1063/1.1703993}
  {\path{doi:http://dx.doi.org/10.1063/1.1703993}}.

\bibitem{Tay06}
B.~A. Tay, G.~Ordonez, Exact Markovian kinetic equation for a quantum Brownian
  oscillator, Phys. Rev. E 73 (2006) 016120.
\newblock \href {http://dx.doi.org/10.1103/PhysRevE.73.016120}
  {\path{doi:10.1103/PhysRevE.73.016120}}.

\bibitem{CL83}
A.~Caldeira, A.~Leggett, Path integral approach to quantum Brownian motion,
  Physica A 121 (1983) 587--616.
\newblock \href {http://dx.doi.org/10.1016/0378-4371(83)90013-4}
  {\path{doi:10.1016/0378-4371(83)90013-4}}.

\bibitem{HPZ92}
B.~L. Hu, J.~P. Paz, Y.~Zhang, Quantum Brownian motion in a general
  environment: Exact master equation with nonlocal dissipation and colored
  noise, Phys. Rev. D 45 (1992) 2843--2861.
\newblock \href {http://dx.doi.org/10.1103/PhysRevD.45.2843}
  {\path{doi:10.1103/PhysRevD.45.2843}}.

\bibitem{Goldstein}
H.~Goldstein, C.~P.~Poole Jr. and J.~L. Safko, Classical Mechanics, 3rd Edition,
  Pearson, Singapore, 2002.

\bibitem{Umezawa75}
Y.~Takahashi, H.~Umezawa, Thermo field dynamics, Coll. Phenom. 2 (1975) 55--80.

\bibitem{Umezawa75b}
Y.~Takahashi, H.~Umezawa, Thermo field dynamics, Int. J. Mod. Phys. B 10 (1996)
  1755--1805.

\bibitem{Tay07}
B.~A. Tay, T.~Petrosky, Thermal symmetry of the Markovian master equation,
  Phys. Rev. A 76 (2007) 042102.
\newblock \href {http://dx.doi.org/10.1103/PhysRevA.76.042102}
  {\path{doi:10.1103/PhysRevA.76.042102}}.

\bibitem{Tay11}
B.~A. Tay, Transformed number states from a generalized Bogoliubov
  transformation and their relationships to the eigenstates of the
  Kossakowski–Lindblad equation, J. Phys. A: Math. Theor. 44 (2011) 255303.
\newblock \href {http://dx.doi.org/10.1088/1751-8113/44/25/255303}
  {\path{doi:10.1088/1751-8113/44/25/255303}}.

\bibitem{Ban93}
M.~Ban, Lie-algebra methods in quantum optics: The Liouville-space formulation,
  Phys. Rev. A 47 (1993) 5093--5119.
\newblock \href {http://dx.doi.org/10.1103/PhysRevA.47.5093}
  {\path{doi:10.1103/PhysRevA.47.5093}}.

\end{thebibliography}
\providecommand{\noopsort}[1]{}\providecommand{\singleletter}[1]{#1}%

\end{document}
%%% 